\documentclass[amstex,11pt,reqno]{article}
\usepackage{amsmath,amsfonts,amssymb,amsthm,enumerate,hyperref,multicol,color,graphicx,epstopdf}
\date{}
\begin{document}
\title{Numerical simulation of MHD boundary layer flow and heat transfer over a nonlinear stretching sheet in the porous medium with viscous dissipation using hybrid approach}
\maketitle
\begin{center}
{\bf Rama Bhargava$^1$, Harish Chandra$^2$}
\vskip0.2in
$^{1,2}$Department of Mathematics\\
Indian Institute of Technology Roorkee\\
Roorkee-247667, India\\
$^1$rbharfma@iitr.ac.in,
$^2$harishchandraamu@gmail.com
\vskip0.3in

\vskip0.1in
\end{center}
\vspace*{0.5cm}

\hspace{-0.5cm}\textbf{Abstract}\\
In the present study, MHD boundary layer flow with heat and mass transfer of a nanofluid with viscous dissipation over a non-linear stretching sheet embedded in a porous medium is studied. The governing non-linear partial differential equations are converted into coupled non-linear ordinary differential equations which are solved by hybrid technique consisting of finite element method with symbolic computation. The effect of Viscous dissipation, magnetic influence parameter, permeability parameter, Brownian motion, Eckert number, Thermophoresis parameter, Prandtl number parameter, Dufour Lewis number and nanofluid Lewis number are studied for velocity $(f'(\eta))$, temperature $(\theta(\eta))$, concentration of salt ($S(\eta)$) and concentration of nanoparticles $(\gamma(\eta))$ and results are shown graphically. Applications of such type of problems are found in the electromagnetic control of complex magnetic nanofluid materials relevant to energy and biomedical systems.\\
\begin{flushleft}
\end{flushleft}
\begin{table}
\begin{tabular}{l l }
{\bf List of Symbols}          &          \\
\bf{Roman Symbols}             &          \\
$K_{m}$                        &     Thermal conductivity\\
$T$                            &     Temperature\\
$T_{w}$                        &     Temperature at the plate
\end{tabular}
\end{table}
\begin{table}
\begin{tabular}{l l }
$T_{\infty}$                   &     Ambient temperature\\
$C$                            &     Solutal concentration\\
$C_{w}$                        &     Solutal concentration at the plate\\
$C_{\infty}$                   &     Ambient solutal concentration\\
$n$                            &     Stretching parameter\\
$D_{B}$                        &     Brownian diffusion coefficient\\
$D_{T}$                        &     Thermophoresis diffusion coefficient\\
$D_{S}$                        &     Solutal diffusivity \\
$D_{TC}$                       &     Dufour diffusivity coefficient\\
$D_{CT}$                       &     Soret diffusivity coefficient\\
\bf{Greek Symbols}             &     \\
$\psi (x,y)$                   &     Stream function\\
$\phi$                         &     Nanoparticle volume fraction\\
$\phi_{w}$                     &     Nanoparticle volume fraction at the plate\\
$\phi_{\infty}$                &     Ambient nanoparticle volume fraction\\
$\nu$                          &     Kinematic viscosity\\
$\rho$                         &     Density\\
$\mu$                          &     Dynamic viscosity\\
$(\rho c)_{p}$                 &     Heat capacity of the nanoparticle\\
$(\rho c)_{f}$                 &     Heat capacity of the fluid\\
$\bf{Subscripts}$              &     \\
$w$                            &     Condition at the surface\\
$\infty$                       &     Condition in the free stream
\end{tabular}
\end{table}
\section{Introduction}
The boundary layer flow over a stretching surface is found to have many industrial and engineering applications. The study of laminar moving boundary layer flow was initiated by Sakiadis\cite{SBC}. Crane\cite{CLJ} extended the flow caused by the stretching sheet while Erickson et al.\cite{EFF} investigated the work of Sakiadis to include suction at the stretching sheet on a continuous surface under constant speed and analyzed its effects on the heat and mass transfer in the boundary layer. Later, many researchers eg. Gupta and Gupta\cite{GPGA}, Dutta et al.\cite{DRG} and Chem et al.\cite{CCCM} continued the work of Crane by considering the effect of heat and mass transfer analysis under different type of physical phenomenon. Mehta et al.\cite{MS} and Gary et al.\cite{GKT} studied the effect when this included the flow characteristics which may change substantially compared to constant viscosity. Reptis\cite{RA}, Vajarvelu and Nayfeh\cite{VN}, Vajarvelu\cite{VK} studied the heat transfer over a stretching sheet with variable viscosity. The flow of heat transfer in both viscous and elastic type of fluid characteristics over a stretching sheet with power law of surface temperature was investigated by Vajravelu and Roper\cite{VR} while Alinejad et al.\cite{AS} investigated this phenomenon in a viscous fluid over a non-linear stretching sheet by including the effects of heat dissipation. Zaimi et al. \cite{ZIP} investigated the fluid flow due to a permeable stretching sheet with viscous effect. Later Dhanai et al. \cite{DPK} extended the work of Zaimi by including the MHD flow with Viscous dissipation effect. Mobood et al. \cite{MKI} also studied the boundary layer flow with heat and mass of an electrically conductively water based nanofluid of a nonlinear stretching sheet with viscous dissipation effect. Mukhopadhayay et al. \cite{MMG} investigated the boundary layer flow of an electrically conducting liquid having variable free stream temperature past a porous stretching sheet in the presence of MHD flow in the porous medium. Jat et al. \cite{JCR} extended this work by including the viscous effect over a nonlinear stretching sheet in the presence of porous medium. Mania et al. \cite{GB} studied the triple diffusive boundary layer flow of nanofluid over a nonlinear stretching sheet.\\
Later Shen et al.\cite{SWC}, Chaudhary and Jat\cite{JRN} investigated the MHD boundary layer flow with heat transfer for stagnation point over a stretching sheet with and without viscous dissipation and Joule heating. Dessie and Kishan\cite{DK} considered the boundary layer flow with heat transfer over a stretching sheet embedded in a porous medium by taking the effects of viscous dissipation and heat source/sink in the presence of uniform magnetic field. Ali et al.\cite{AAA} studied the hydro-magnetic flow and heat transfer of fluid over an inclined stretching plate.\\
The mathematical model of nanofluids was first proposed by Buongiorno\cite{BJ}. He concluded that there are seven slip mechanisms which are responsible for convective transport in nanofluids out of which brownian motion, thermophoresis are the most important. Other models for nanofluids were proposed by Khanafer et al.\cite{KV}. Kuznetsov and Nield {\cite{KN,KAN}} utilized the mathematical model of nanofluid suggested by Buongiorno\cite{BJ} to study some problems on viscous fluids and porous media filled by nanofluids. Rana and Bhargava\cite{RB} investigated the boundary layer flow with heat transfer of a nanofluid over a non-linearly stretching sheet. The problem of MHD with nanofluid convection was studied by yadav et al.\cite{YWBL,YL}.\\
The purpose of this work is to study the viscous flow and heat transfer over a nonlinear stretching sheet embedded in a porous medium including the MHD flow in the presence of nanofluid with viscous dissipation. This work also deals with the triple diffusive boundary layer flow. Using the similarity transformation, the governing partial differential equations are converted into coupled nonlinear ordinary differential equations, which are solved numerically by hybrid approach consisting of finite element method with symbolic computation. The comparison of results and computational time for the different schemes are presented.\\
Symbolic computation is the study of algorithms for handling mathematical expressions which focuses attention on exact computing of expressions containing variables. The variables are not given any values and are interpreted and manipulated as symbols, and thus, the name symbolic computation. Felippa\cite{FCL} discussed a set of Mathematica modules of numerical integration rules, which are then used for symbolic finite element work. Thus significant time savings has been established.
Videla\cite{VGCB} studied the accurate Integration of the Stiffness Matrix of an 8-Node Plane Elastic Finite Element by Symbolic Computation. A nice review of symbolic computation with FEM has been given by Pillwein\cite{PV}. One of the important aim of the current study is to justify the computational efficiency of the hybrid technique.
\section{Mathematical Formulation}
Consider the two dimensional boundary layer flow of an incompressible viscous and nanofluid past a non-linear stretching sheet with velocity $u(x)=c x^{n}$, where $c>0$ is a constant and $n>1$ is the non-linear stretching parameter. The x-coordinate is measured along the non linear stretching sheet and y-coordinate along perpendicular to the surface. A uniform magnetic field of strength $B_{0}$ is applied perpendicular to the non-linear stretching sheet. The magnetic Reynolds number is assumed to be small and therefore the induced magnetic field is neglected. It is assumed that the constant temperature, solutal and nanoparticle volume fraction at the surface of the non linear stretching sheet are $T_{w}$, $C_{w}$ and $\phi_{w}$ respectively. These values are assumed to be greater than the ambient temperature ($T_{\infty}$), solutal concentration ($C_{\infty}$) and nanoparticle concentration ($\phi_{\infty}$) away from the sheet. This is shown in figure 1.
\begin{center}
\includegraphics[width=0.75\columnwidth]{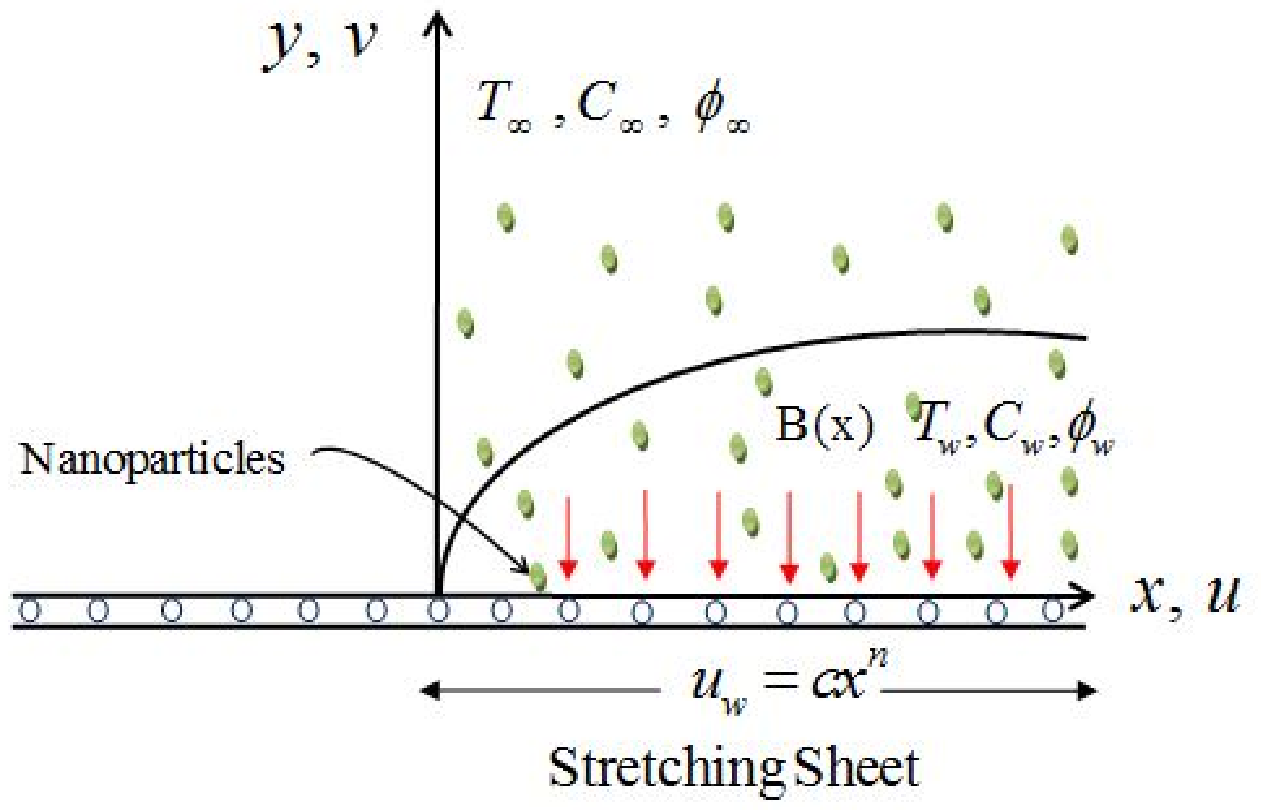}
\end{center}
\begin{center}
\textbf{Figure 1: Physical model and coordinate system of the problem.}
\end{center}
Under the above assumptions, the five governing equations of the conservation of mass, conservation of momentum, conservation of energy, concentration of salt and concentration of nanoparticle for nanofluid can be written as follows:
\begin{align}
&\frac{\partial u}{\partial x}+\frac{\partial v}{\partial y}=0\label{e1}\\
&u\frac{\partial u}{\partial x}+v\frac{\partial u}{\partial y}=\nu\frac{\partial^2{u}}{\partial y^2}-\frac{\sigma B^{2}(x)}{\rho}u-\frac{\nu}{k(x)}u\label{e2}\\
&u\frac{\partial T}{\partial x}+v\frac{\partial T}{\partial y}=\alpha_{m}\frac{\partial^2{T}}{\partial y^2}+\frac{\mu}{\rho c_{P}}\left(\frac{\partial u}{\partial y}\right)^2+\tau \left ( D_B \frac{\partial \phi}{\partial y}\frac{\partial T }{\partial y}+\frac{D_T}{T_\infty}\left({\frac{\partial T}{\partial y}}\right)^2 \right)+D_{TC}\frac{\partial^{2} C}{\partial y^{2}}\label{e3}\\
&u \frac{\partial C}{\partial x}+v\frac{\partial C}{\partial y}= D_{S}\frac{\partial^2 C}{\partial y^2}+D_{CT} \frac{\partial^2 T}{\partial y^2}\label{e4}\\
&u \frac{\partial \phi}{\partial x}+v\frac{\partial \phi}{\partial y}= D_B\frac{\partial^2 \phi}{\partial y^2}+\frac {D_T}{T_\infty}\frac{\partial^2 T}{\partial y^2}\label{e5}
\end{align}
where\\
$\alpha_{m}=\displaystyle\frac{K_{m}}{(\rho c)_{f}}$,  $\tau=\displaystyle\frac{(\rho c)_{P}}{(\rho c)_{f}}$.\\
and ($u,v$) are the velocity components along the x-axis and y-axis respectively, $\alpha_{m}$ is the thermal diffusivity, $\sigma$ is electrical conductivity of the fluid, and we assume that the variable magnetic field B(x) and variable permeability k(x) are in the form of $B(x)=B_{0}x^{(n-1)/2}$ and $k(x)=k_{0}x^{(1-x)}$ respectively \cite{MKI,MMG}. $\tau$ is defined as the ratio between the effective heat capacity of the nanoparticles and heat capacity of the fluid with $c_{p}$ as the specific heat at constant pressure. In the equation (\ref{e3}) on the RHS second term denotes the viscous dissipation and fourth term represents the energy flux \cite{GB}. The boundary conditions are,
\begin{align}\label{e6}
&u= c x^n,~~ v=0,~~ T=T_{w},~~C=C_{w},~~ \phi=\phi_{w}~~ at ~~y=0\nonumber\\
&u\rightarrow 0,~~v\rightarrow0,~~T\rightarrow T_{\infty},~~C\rightarrow C_{\infty},~~ \phi\rightarrow \phi_{\infty}~~as~~y\rightarrow \infty.
\end{align}
Considering the stream function $\psi$ such that
\begin{align}\label{e7}
u=\frac{\partial \psi}{\partial y},~~v=-\frac{\partial \psi}{\partial x}.
\end{align}
 and introducing the following similarity transformations,
\begin{align}\label{e8}
&\eta=  y\sqrt{\frac{c(n+1)}{2\nu}} x^{(n-1)/2},~~u=c x^{n} f'(\eta),~~v=-\sqrt{\frac{c\nu(n+1)}{2}}x^{(n-1)/2}\left[f+\frac{n-1}{n+1}\eta f'\right]\nonumber\\
&\theta(\eta)=\frac{T-T_{\infty}}{T_{w}-T_{\infty}},~~S(\eta)=\frac{C-C_{\infty}}{C_{w}-C_{\infty}},~~
\gamma(\eta)=\frac{\phi-\phi_{\infty}}{\phi_{w}-\phi_{\infty}}
\end{align}
the governing non-linear equations (\ref{e1})-(\ref{e5}) reduce to the following set of non-linear ordinary differential equations:
\begin{align}
&f'''-\left(\frac{2 n}{n+1}\right)f'^{2}+f f''- (M+R) f'=0\label{e9}\\
&\theta''+Pr \left[f \theta'+Ec f''^{2}+Nt \theta'^{2}+Nb S' \theta'+ {Nd} \gamma''\right]=0\label{e10}\\
&S''+Le f S'+ {Ld}\theta''=0\label{e11}\\
&\gamma''+Ln f \gamma'+\frac{Nt}{Nb}\theta''=0\label{e12}
\end{align}
with the corresponding boundary conditions (\ref{e6}) as
\begin{align}\label{e13}
&f(\eta)=0, f'(\eta)=1, \theta(\eta)=1, S(\eta)=1, \gamma(\eta)=1~~ at~~ \eta=0\nonumber\\
&           f'(\eta)\rightarrow 0, \theta(\eta)\rightarrow 0, S(\eta)\rightarrow 0, \gamma(\eta)\rightarrow 0~~as ~~\eta\rightarrow \infty
\end{align}
Dimensionless parameters used in equations (\ref{e9})-(\ref{e12}) are:\\
$M=\displaystyle\frac{2 \sigma B^{2}_{0}}{c \rho_{f} (n+1)}$ is the magnetic parameter, $R=\displaystyle\frac{2\nu}{ck_{0}(n+1)}$ is the permeability parameter, $Pr=\displaystyle\frac{\nu}{\alpha}$ is the Prandtl number and $Ec=\displaystyle\frac{u_{w}^2}{c_{p}\Delta T}$ is the Eckert number, $Nb=\displaystyle\frac{D_{B}(\rho c)_{P} (\phi_{w}-\phi_{\infty})}{(\rho c)_{f}\nu}$ is the Brownian motion parameter, $Nt=\displaystyle\frac{(\rho c)_{P}D_{T}\Delta T}{(\rho c)_{f}\nu T_{\infty}}$ is the thermophoresis parameter, $Nd=\displaystyle \frac{D_{TC}(C_{w}-C_{\infty})}{\nu(T_{w}-T_{\infty})}$ is the modified Dufour parameter of salt, $Ld=\displaystyle\frac{D_{TC}(T_{w}-T_{\infty})}{D_{S}(C_{w}-C_{\infty})}$ is the Dufour-solutal Lewis number, $ Le=\displaystyle\frac{\nu}{D_{S}}$ is the regular Lewis number of salt, $Ln=\displaystyle\frac{\nu}{D_{B}}$ is the nanofluid Lewis number. Expressions for the local skin friction coefficient ($c_{f}$), the Nusselt number ($Nu_{x}$), the Sherwood number ($Sh_{x}$) and the nanoparticle Sherwood number ($Sh_{x,n}$) are,\\
\begin{align}\label{e14}
c_{f}=\displaystyle \frac{\mu_{f}}{\rho u^{2}_{w}} \left(\frac{\partial u}{\partial y}\right)_{y=0},~
Nu_{x}=\displaystyle\frac{x q_{w}}{k_{m}\left(T_{w}-T_{\infty}\right)},~Sh_{x}=\displaystyle \frac{x q_{m}}{D_{S}\left(C_{w}-C_{\infty}\right)},~ Sh_{x,n}=\displaystyle \frac{x q_{np}}{D_{B}\left(\phi_{w}-\phi_{\infty}\right)}
\end{align}
where $q_{w}$ is the heat flux, $q_{m}$  and $q_{np}$ are the regular and nanofluid mass flux at the surface of the sheet is defined as.\\
\begin{align}\label{e15}
q_w= \displaystyle -K_{m} \left(\frac{\partial T}{\partial y}\right)_{y=0},~~ q_{m}= \displaystyle -D_{S} \left(\frac{\partial C}{\partial y}\right)_{y=0},~~ q_{np}= \displaystyle -D_{B} \left(\frac{\partial \phi}{\partial y}\right)_{y=0}.
\end{align}
 Substituting the values from equation (\ref{e8}) in equations (\ref{e14})-(\ref{e15}), we get\\
\begin{align}\label{e16}
Re^{1/2}_{x}C_{f}=\displaystyle \sqrt{\frac{(n+1)}{2 }}f''(0),~~
\displaystyle \frac{Nu_{x}}{\sqrt{\frac{n+1}{2}Re_{x}}}=-\theta'(0)=Nur, \nonumber \\
\displaystyle \frac{Sh_{x}}{\sqrt{\frac{n+1}{2}Re_{x}}}=-S'(0)=Shr,~~
\displaystyle \frac{Sh_{x,n}}{\sqrt{\frac{n+1}{2}Re_{x}}}=-\gamma'(0)=Shrn
\end{align}
where $Nur$, $Shr$ and $Shrn$ are the reduced form of Nusselt, Sherwood number and nanoparticle Sherwood number respectively, and the local Reynolds number based on the stretching velocity is $Re_{x}=\displaystyle \frac{u_{x}x}{\nu}$.
The ordinary differential equations (\ref{e9})-(\ref{e12}) are highly non-linear, which are solved numerically by Hybrid technique.
\section{Numerical Solution}
The latest hybrid technique which consists of the FEM with symbolic computation is used for solving the present problem. The error is minimized using hybrid technique which is inherent in FEM due to numerical integration. Thus this approach is expected to yield better and efficient results. The steps used in the hybrid technique are:\\
(i) Discretization of the domain into elements. \\
(ii) Generation of the element equations with symbolic computation. \\
(iii) Assembly of the element equations. \\
(iv) Imposition of the boundary equations.\\
(v) Solution of the assembled equations.\\
By Assuming
\begin{align}\label{e17}
 &f'=g
 \end{align}
 The simultaneous differential equations (\ref{e9})-(\ref{e12}) subject to the boundary conditions (\ref{e13}) are reduced in lower order as follows:
\begin{align}
&g''-\left(\frac{2 n}{n+1}\right)g^{2}+f g'-(M+R)g=0\label{e18}\\
&\theta''+Pr \left(f \theta'+Ec g'^{2}+Nt \theta'^{2}+Nb \gamma' \theta'+Nd S''\right)=0\label{e19}\\
&S''+Le f S'+Ld\theta''=0\label{e20}\\
&\gamma''+Ln f \gamma'+\frac{Nt}{Nb}\theta''=0\label{e21}
\end{align}
and the appropriate boundary conditions are
\begin{align}\label{e22}
&f(\eta)=0, g(\eta)=1, \theta(\eta)=1, S(\eta)=1, \gamma(\eta)=1 ~~ at~~ \eta=0\nonumber\\
&           g(\eta)\rightarrow 0, \theta(\eta)\rightarrow 0, S(\eta)\rightarrow 0, \gamma(\eta)\rightarrow 0 ~~as ~~\eta\rightarrow \infty
\end{align}
\subsection{ Variational Formulation}
The variational form associated with equations (\ref{e17})-(\ref{e21}) over the element ($\eta_e,  \eta_{e+1}$) can be written as
\begin{align}
& \int_{\eta_e}^{\eta_{e+1}} w_1\left(f'-g \right)\,d\eta=0 \label{e23}\\
& \int_{\eta_e}^{\eta_{e+1}} w_2\left(g''-\left(\frac{2 n}{n+1}\right)g^{2}+f g'-(M+R)g\right)\,d\eta=0\label{e24}\\
& \int_{\eta_e}^{\eta_{e+1}} w_3\left(\theta''+Pr \left(f \theta'+Ec g'^{2}+Nt \theta'^{2}+Nb \gamma' \theta'+Nd S''\right)\right)\,d\eta=0\label{e25} \\
&\int_{\eta_e}^{\eta_{e+1}} w_4\left(S''+Le f S'+Ld\theta''\right)\,d\eta=0\label{e26}\\
&\int_{\eta_e}^{\eta_{e+1}} w_4\left(\gamma''+Ln f \gamma'+\frac{Nt}{Nb}\theta''\right)\,d\eta=0\label{e27}
\end{align}
where $w_1$, $w_2$, $w_3$, $w_4$ and $w_5$ are arbitrary weight functions and may be regarded as the variation in $f$, $g$, $\theta$, $S$ and $\gamma$ respectively and domain ($\eta_e, \eta_{e+1}$) denotes the length of the boundary layer region.
\subsection{Finite element formulation}
Let the domain be defined into three-noded quadratic elements.The corresponding finite element approximation are:
\begin{align}\label{e28}
& f=\sum_{j=1}^{3}f_j \psi_j,~~~g=\sum_{j=1}^{3}g_j \psi_j,~~~\theta=\sum_{j=1}^{3}\theta_j \psi_j,\nonumber\\
&S=\sum_{j=1}^{3}S_{j} \psi_j,~~~\gamma=\sum_{j=1}^{3}\gamma_j \psi_j
\end{align}
with $w_1=w_2=w_3=w_4=w_5=\psi_i$, ($i=1,2,3$). The quadratic shape functions $\psi_i$ are:
\begin{align}\label{e29}
&\psi_1^e=\frac{(\eta_{e+1}-\eta)(\eta_{e+1}+\eta_{e}-2\eta)}{(\eta_{e+1}-\eta_e)^2},~~~~\psi_2^e=\frac{4(\eta-\eta_{e})(\eta_{e+1}-\eta)}{(\eta_{e+1}-\eta_e)^2},\nonumber\\
&\psi_3^e=-\frac{(\eta-\eta_e)(\eta_{e+1}+\eta_{e}-2\eta)}{(\eta_{e+1}-\eta_e)^2}\end{align}
where $\displaystyle\eta~\epsilon~[\eta_{e}~~\eta_{e+1}]$. Thus the finite-element model of these equations are given by
\begin{align}\label{e30}
&\begin{bmatrix} [k^{11}] & [k^{12}] & [k^{13}] & [k^{14}] & [k^{15}] \\ [k^{21}] & [k^{22}] & [k^{23}] & [k^{24}] & [k^{25}]  \\ [k^{31}] & [k^{32}] & [k^{33}] & [k^{34}] & [k^{35}] \\ [k^{41}] & [k^{42}] & [k^{43}] & [k^{44}] & [k^{45}] \\ [k^{51}] & [k^{52}] & [k^{53}] & [k^{54}] & [k^{55}] \end{bmatrix} \left[ \begin{array}{c} f \\ g \\ \theta \\ S \\ \gamma \end{array} \right]=\left[ \begin{array}{c} \{h^1\} \\ \{h^2\} \\ \{h^3\} \\ \{h^4\} \\ \{h^5\} \end{array} \right]
\end{align}
Here each [$k^{mn}$] is of the order $3\times 3$ and [$h^{m}$], ($m,n = 1,2,3,4,5$) is of $3\times 1$. These matrices are defined as:
\begin{align}
& k^{11}_{ij}=\int_{\eta_e}^{\eta_{e+1}} {\psi_i} \frac{\partial \psi_j}{\partial \eta} \,d\eta ,~ k^{12}_{ij}=-\int_{\eta_e}^{\eta_{e+1}} \psi_i \psi_j \,d\eta ,~ k^{13}_{ij}=0,~ k^{14}_{ij}=0,~ k^{15}_{ij}=0 ;\label{e31}\\
& k^{21}_{ij}=0 ,~ k^{22}_{ij}=-\int_{\eta_e}^{\eta_{e+1}} \frac{\partial \psi_i}{\partial \eta} \frac{\partial \psi_j}{\partial \eta} \,d\eta-\left(\frac{2n}{n+1}\right)\int_{\eta_e}^{\eta_{e+1}} \psi_i \bar g \psi_j \,d\eta \nonumber\\
&+\int_{\eta_e}^{\eta_{e+1}} {\psi_i} \bar f \frac{\partial \psi_j}{\partial \eta} \,d\eta-(M+R) \int_{\eta_e}^{\eta_{e+1}} {\psi_i} {\psi_j} \,d\eta,~ k^{23}_{ij}=0,~k^{24}_{ij}=0,~ k^{25}_{ij}=0;\label{e32}\\
& k^{31}_{ij}=0 ,~ k^{32}_{ij}=Pr Ec\int_{\eta_e}^{\eta_{e+1}} {\psi_i} \bar g' \frac{\partial \psi_j}{\partial \eta} \,d\eta , ~k^{33}_{ij}=-\int_{\eta_e}^{\eta_{e+1}} \frac{\partial \psi_i}{\partial \eta} \frac{\partial \psi_j}{\partial \eta} \,d\eta\nonumber\\
&+Pr \int_{\eta_e}^{\eta_{e+1}} {\psi_i} \bar f \frac{\partial \psi_j}{\partial \eta} \,d\eta+Pr Nt\int_{\eta_e}^{\eta_{e+1}} {\psi_i} \bar \theta' \frac{\partial \psi_j}{\partial \eta} \,d\eta+Pr Nb\int_{\eta_e}^{\eta_{e+1}} {\psi_i} \bar \gamma' \frac{\partial \psi_j}{\partial \eta} \,d\eta\nonumber\\ &k^{34}_{ij}=-Pr Nd \int_{\eta_e}^{\eta_{e+1}} \frac{\partial \psi_i}{\partial \eta} \frac{\partial \psi_j}{\partial \eta} \,d\eta,~ k^{35}_{ij}=0;\label{e33}\\
& k^{41}_{ij}=0 ,~ k^{42}_{ij}=0 ,~ k^{43}_{ij}=-Ld \int_{\eta_e}^{\eta_{e+1}} \frac{\partial \psi_i}{\partial \eta} \frac{\partial \psi_j}{\partial \eta} \,d\eta ,~k^{44}_{ij}=- \int_{\eta_e}^{\eta_{e+1}} \frac{\partial \psi_i}{\partial \eta} \frac{\partial \psi_j}{\partial \eta} \,d\eta \nonumber\\
&+Le \int_{\eta_e}^{\eta_{e+1}} \psi_i \bar f \frac{\partial \psi_j}{\partial \eta} \,d\eta ,~ k^{45}_{ij}=0;\label{e34}\\
& k^{51}_{ij}=0 ,~ k^{52}_{ij}=0 ,~ k^{53}_{ij}=-\frac{Nt}{Nb} \int_{\eta_e}^{\eta_{e+1}} \frac{\partial \psi_i}{\partial \eta} \frac{\partial \psi_j}{\partial \eta} \,d\eta,~k^{54}_{ij}=0,~k^{55}_{ij}= -\int_{\eta_e}^{\eta_{e+1}} \frac{\partial \psi_i}{\partial \eta} \frac{\partial \psi_j}{\partial \eta} \,d\eta \nonumber\\
&+Ln \int_{\eta_e}^{\eta_{e+1}} \psi_i \bar f \frac{\partial \psi_j}{\partial \eta} \,d\eta ;\label{e35}
\end{align}
\begin{align}
&h^1_i=0 ,~ h^2_i=-\left(\psi_i\frac{\partial g}{\partial \eta}\right)^{\eta_{e+1}}_{\eta_e} ,~ h^3_i=-\left(\psi_i\frac{\partial \theta}{\partial \eta}\right)^{\eta_{e+1}}_{\eta_e} ,~h^4_i=-\left( Nd \frac{\partial \theta}{\partial \eta} + \frac{\partial S}{\partial \eta}\right)^{\eta_{e+1}}_{\eta_e}\nonumber\\
&h^5_i=-\left(\frac{Nt}{Nb} \frac{\partial \theta}{\partial \eta} + \frac{\partial \gamma}{\partial \eta}\right)^{\eta_{e+1}}_{\eta_e};\label{e36}
\end{align}
where
\begin{align}\label{e37}
\bar f'=\sum_{j=1}^{3}\bar f_j \frac{\partial\psi_j}{\partial\eta},~~~\bar g'=\sum_{j=1}^{3}\bar g_j \frac{\partial\psi_j}{\partial\eta},~~~\bar\theta'=\sum_{j=1}^{3}\bar\theta_j \frac{\partial\psi_j}{\partial\eta},~~~~\bar S'=\sum_{j=1}^{3}\bar S_j \frac{\partial\psi_j}{\partial\eta},~~~\bar \gamma'=\sum_{j=1}^{3}\bar \gamma_j \frac{\partial\psi_j}{\partial\eta}
\end{align}
\subsection{Symbolic Computation}
All integrations are computed symbolical by with the help of mathematica 10. Thus,
\begin{align}
&\int_{0}^{h_e} \psi_i \psi_j \,d\bar x=\begin{bmatrix} (2 h_e)/15 & h_e/15 & -(h_e/30)\\h_e/15 & (8 h_e)/15& h_e/15\\ -(h_e/30)& h_e/15 & (2 h_e)/15 \end{bmatrix}=k_{111} ,\nonumber\\
&\int_{0}^{h_e} {\psi_i} \frac{\partial \psi_j}{\partial \bar x} \,d\bar x=\begin{bmatrix} -(1/2) & 2/3 & -(1/6) \\ -(2/3) & 0 & 2/3 \\ 1/6 & -(2/3) & 1/2 \end{bmatrix}=k_{112} ,\nonumber
\end{align}
\begin{align}
&\int_{0}^{h_e} \frac{\partial \psi_i}{\partial \bar x} \frac{\partial \psi_j}{\partial \bar x} \,d\bar x=\begin{bmatrix} 7/(3 h_e) & -8/(3 h_e) & 1/(3 h_e) \\ -8/(3 h_e) & 16/(3 h_e) & -8/(3 h_e) \\ 1/(3 h_e) & -8/(3 h_e) & 7/(3 h_e) \end{bmatrix}=k_{113} ;\label{e38}\\
&\int_{0}^{h_e} \psi_i\psi_1 \psi_j \,d\bar x=\begin{bmatrix} 13 h_e/140 & h_e/21 & -h_e/140 \\ h_e/21 & 4 h_e/105 & -2 h_e/105 \\ -h_e/140 & -2 h_e/105 & -h_e/140 \end{bmatrix}=k_{221} ,\nonumber\\
&\int_{0}^{h_e} \psi_i\psi_2 \psi_j \,d\bar x=\begin{bmatrix} h_e/21 & 4 h_e/105 & -2 h_e/105\\ 4 h_e/105 & 16 h_e/35 & 4 h_e/105 \\ -2 h_e/105 & 4 h_e/105 & h_e/21 \end{bmatrix}=k_{222} ,\nonumber\\
&\int_{0}^{h_e} \psi_i\psi_3 \psi_j \,d\bar x=\begin{bmatrix} -h_e/140 & -2 h_e/105 & -h_e/140 \\ -2 h_e/105 & 4 h_e/105 & h_e/21 \\ -h_e/140 & h_e/21 & 13 h_e/140 \end{bmatrix}=k_{223} ;\label{e39}\\
&\int_{0}^{h_e} {\psi_i}\psi_1 \frac{\partial \psi_j}{\partial \bar x} \,d\bar x= \begin{bmatrix} -1/3 & 2/5 & -1/15 \\ -1/5 & 4/15 & -1/15 \\ 1/30 & 0 & -1/30 \end{bmatrix}=k_{331},\nonumber\\
&\int_{0}^{h_e} {\psi_i}\psi_2 \frac{\partial \psi_j}{\partial \bar x} \,d\bar x=\begin{bmatrix} -1/5 & 4/15 & -1/15 \\ -8/15 & 0 & 8/15 \\ 1/15 & -4/15 & 1/5 \end{bmatrix}=k_{332} ,\nonumber\\
&\int_{0}^{h_e} {\psi_i}\psi_3 \frac{\partial \psi_j}{\partial \bar x} \,d\bar x=\begin{bmatrix} 1/30 & 0 & -1/30 \\ 1/15 & -4/15 & 1/5 \\ 1/15 & 2/5 & 1/3  \end{bmatrix}=k_{333} ;\label{e40}\\
&\int_{0}^{h_e} \frac{\partial \psi_i}{\partial \bar x}\psi_1 \frac{\partial \psi_j}{\partial \bar x} \,d\bar x=\begin{bmatrix} 37/30 h_e & -22/15 h_e & 7/30 h_e \\ -22/15 h_e & 8/5 h_e & -2/15 h_e \\ 7/30 h_e & -2/15 h_e & -1/10 h_e \end{bmatrix}=k_{441} ,\nonumber\\
&\int_{0}^{h_e} \frac{\partial \psi_i}{\partial \bar x}\psi_2 \frac{\partial \psi_j}{\partial \bar x} \,d\bar x=\begin{bmatrix} 6/5 h_e & -16/15 h_e & -2/15 h_e \\ -16/15 h_e & 32/15 h_e & -16/15 h_e \\ -2/15 h_e & -16/15 h_e & 6/5 h_e \end{bmatrix}=k_{442} ,\nonumber\\
&\int_{0}^{h_e} \frac{\partial \psi_i}{\partial \bar x}\psi_3 \frac{\partial \psi_j}{\partial \bar x} \,d\bar x=\begin{bmatrix} -1/10 h_e & -2/15 h_e & 7/30 h_e \\ -2/15 h_e & 8/5 h_e & -22/15 h_e \\ 7/30 h_e & -22/15 h_e & 37/30 h_e \end{bmatrix}=k_{443};\label{e41}
\end{align}
Using these integral values in equations (\ref{e31})-(\ref{e35}), we have
\begin{align}
&k^{11}=k_{112},~k^{12}=-k_{111},~k^{13}=0,~k^{14}=0,~k^{15}=0 ;\label{e42}\\
& k^{21}=0 ,~k^{22}=-k_{113}-\left(\frac{2n}{n+1}\right)(k_{221}g(1)+k_{222} g(2)+k_{223} g(3))+(k_{331} f(1)\nonumber\\
&+k_{332} f(2)+k_{333} f(3))-(M+R) k_{111},~k^{23}=0,~k^{24}=0,~k^{25}=0 ;\label{e43}\\
&k^{31}=0 ,~k^{32}=(Pr) (Ec) (k_{441}g(1)+k_{442} g(2)+k_{443} g(3)) ,k^{33}=-k_{113}+(Pr) (k_{331}f(1)\nonumber\\
&+k_{332} f(2)+k_{333} f(3))+(Pr)(Nt) (k_{441} \theta(1)+k_{442} \theta(2)+k_{443} \theta(3))\nonumber\\
&+(Pr)(Nb)(k_{441} \gamma(1)+k_{442} \gamma(2)+k_{443} \gamma(3)),~k^{34}=-(Pr) (Nd) k_{113},~k^{35}=0 ;\label{e44}
\end{align}
\begin{align}
&k^{41}=0 ,~k^{42}=0 ,~k^{43}=-(Ld) k_{113} ,~k^{44}= - k_{113}+(Le) (k_{331}f(1)+k_{332} f(2)\nonumber\\
&+k_{333} f(3)),~k^{45}=0;\label{e45}\\
&k^{51}=0 ,~k^{52}=0 ,~k^{53}=-\left(\frac{Nt}{Nb}\right) k_{113} ,~k^{54}=0 ,~k^{55}=- k_{113}+(Ln) (k_{331}f(1)\nonumber\\
&+k_{332} f(2)+k_{333} f(3))\label{e46} ;
\end{align}
and $h^1,~~h^2,~~h^3,~~h^4,~~h^5$ are also shown in equation (\ref{e36}).\\
\textbf{Table 1.}\\
Grid-independency test for velocity gradient, for different parameters $Pr=0.5, Nt=0.2, Nb=0.1, M=0.5, Le=2, Ec=0.4, n=2, Nd=0.2, Ln=2, Ld=0.1, R=0.5$
\begin{center}
 \begin{tabular}{|l|lll|}
 \hline
                            &                 &        $f''(0)$  &                 \\
 \cline{2-4}
 step  &                 &                  &                 \\
 size  & $\eta_\infty$=4 & $\eta_\infty$=6  & $\eta_\infty$=8 \\
 \hline
 0.08  &1.003519 &0.995904 &0.989415\\
 0.04  &1.009935 &1.005531 &1.002272\\
 0.02  &1.013137 &1.010332 &1.008679\\
 0.01  &1.014737 &1.012730 &1.011876\\
 0.005 &1.015536 &1.013927 &1.013473\\
 0.004 &1.015696 &1.014167 &1.013793\\
 \hline
 \end{tabular}
\end{center}
Each element matrix is of order $15\times15$. The computation domain is discretized with uniformly distributed 1200 quadratic elements and following assembly, the assembled matrix of order $6005\times6005$ is generated. The assembled system of equations is non-linear and thus, the iterative scheme is applied to solve it. The system is linearized by incorporating the functions $\bar f$, $\bar g$, $\bar\theta$, $\bar S$ and $\bar \gamma$, which are assumed to be known. The iterative process is terminated when the following condition is satisfied
$$\sum_{i,j}{|\Phi_{i,j}^{m^{*}}-\Phi_{i,j}^{m^{*}-1}|\leq 10^{-4}}$$
where $\Phi$ stands for either $f$, $g$, $\theta$, $S$ and $\gamma$, and $m^{*}$ denotes the iterative step. Excellent convergence has been achieved for all the results. By grid independency test (Table 1), it was concluded that $\eta_\infty=6$ ensures grid independent solutions.\\

\section{Results and Discussion}
In order to study the physical behaviour of velocity, temperature, solutal concentration and nanoparticle concentration functions, a comprehensive numerical computations have been carried out for different values of the parameters that describe the flow characteristics and the results are presented both graphically and in tabular form. In all the cases, the following default values of the parameters are taken as $Pr=0.5,~~Nt=0.2,~~Nb=0.1$, $M=0.5,~~Le=2,~~Ec=0.4,~~n=2$, $Ln=2$, $Nd=0.2$, $Ld=0.1,~~R=0.5$. These values are selected in accordance with the earlier studies, as available in literature. The numerical method is validated with the result obtained by Cortell \cite{CR} in terms of reduced nusselt number and the comparison is illustrated in table 2. Computational time for both finite element method and hybrid approach are compared for different number of elements and it is shown in the table 3. From this table it is observed that the computational time for hybrid technique is much lesser than the finite element method. \textbf{Which shows the efficiency of the Hybrid scheme. Thus is solving the complicated problems by Hybrid approach will defiantly bring a grate cut off in the computational time, bring here efficiency. This was the basic objective behaved this work}.\\
\textbf{Table 2.}\\
Reduced Nusselt number ($Nur$) comparison for with Cortell \cite{CR} for $M=R=Nb=Nt=Nd=0$.
\begin{center}
\begin{tabular}{|l|l|ll|llll|}
\hline
      &   &        &               &&Present Results&                    &\\
\cline{5-8}
Ec    & n & Results in& Cortell \cite{CR}&                 Using FEM     &&Using Hybrid        &\\
      &   &   &&              &                & Approach            &\\
      &   &                &               &                &        &            &\\
\cline{3-8}
      &    & Pr=1    & Pr=5    &  Pr=1   &  Pr=5  & Pr=1   & Pr=5   \\
\hline
      &0.2 &0.610262 &1.607175 &0.667866 &1.756909 &0.668006 &1.757222\\
      &0.5 &0.595277 &1.586744 &0.620472 &1.654177 &0.620684 &1.654647\\
 0.0  &1.5 &0.574537 &1.557463 &0.565300 &1.532037 &0.565573 &1.532631\\
      &3.0 &0.564472 &1.542337 &0.541195 &1.478927 &0.541485 &1.479559\\
      &10  &0.554960 &1.528573 &0.519295 &1.430756 &0.519599 &1.431417\\
      &0.2 &0.574985 &1.474764 &0.633658 &1.460049 &0.633777 &1.460217\\
 0.1  &0.5 &0.556623 &1.436789 &0.592936 &1.401035 &0.593122 &1.401321\\
      &1.5 &0.530966 &1.381861 &0.543852 &1.323561 &0.544096 &1.323959\\
      &3.0 &0.517977 &1.352768 &0.521955 &1.287093 &0.522218 &1.287539\\
      &10  &0.505121 &1.324772 &0.501866 &1.252861 &0.502144 &1.253341\\
\hline
\end{tabular}
\end{center}

\textbf{Table 3.}\\
Comparison of computational time between FEM and Hybrid technique for different number of elements.
\begin{center}
\begin{tabular}{|l|ll|ll|}
\hline
 No. of   &FEM              &                 & Hybrid           &                  \\
 elements &                 &                 & technique        &                  \\
\cline{2-5}
step      &                 &                 &                  &                  \\
size      & $f''(0)$        & Computation time& $f''(0)$         & Computation time \\
          &                 & in seconds      &                  & in seconds       \\
\hline
10        &0.918425         &0.723252         &0.918062          &0.509240           \\
50        &0.996190         &15.597698        &0.995904          &10.631567          \\
100       &1.005808         &58.640368        &1.005531          &40.121631          \\
200       &1.010604         &216.939189       &1.010332          &148.736507         \\
300       &1.012200         &469.432319       &1.011931          &324.002991         \\
500       &1.013477         &1241.277890      &1.013209          &873.272590         \\
1000      &1.014434         &4747.749588      &1.014167          &3431.183689        \\
15000     &1.014753         &10581.821882     &1.014486          &7831.559303        \\
20000     &1.014912         &18926.036666     &1.014646          &14301.883595       \\

\hline
\end{tabular}
\end{center}

Figures 2(a), 2(b) and 2(c) represent the velocity, temperature and nanoparticle concentration profiles respectively for different values of magnetic parameter $M$. It is noticed that the velocity profile in the boundary layer region decreases and temperature along with nanoparticle concentration increases with increasing the value of $M$. The fluid velocity reduced with the increment of magnetic parameter ($M$) which shown in fig. 2(a) that validates the physical behaviour of the magnetic flied. It is clear that the magnetic parameter defines the ratio of Lorentz force to the viscus force. Therefore large value of $M$ indicates that the Lorentz force enhances. This force produces more resistance to the transport phenomena due to which the velocity of fluid reduces. Hence the boundary layer thickness is a depreciating function of $M$. From the figs. 2(b) and 2(c) it is observed that the temperature profile and nanoparticle concentration profile enhances with the increment values of $M$. Since the Lorentz force oppose the fluid motion due to which the temperature is increase the thermal boundary layer as well as fluid temperature. On the other hand, the Lorentz force has the tendency to increase the nanoparticle volume fraction in the motion of the nanofluid. So, the magnetic field increases the nanoparticle volume fraction in the boundary layer regime.\\
Figure 3 illustrates the behaviour of the temperature profile for different values of Eckert number ($Ec$). This figure indicates that the thermal boundary layer thickness increases with the increment of Eckert number ($Ec$). It is obvious that the Eckert number controls the fluid motion and the value of $Ec=0$ represents no viscous dissipation because Eckert number ($Ec$) defines the ratio of the square of the fluid velocity far from the surface boundary to the product of the specific heat of the fluid at constant temperature and the difference between the temperatures of the fluid and the surface boundary. In the case of viscous dissipation ($Ec>0$ or $Ec<0$), the energy equation acts as an internal energy due to the action of viscous stress, so that the thermal boundary layer increases with the value of $Ec$.\\
Figures 4(a) and 4(b) depict the effect of non-linear stretching parameter ($n$) on the velocity and temperature profiles through the boundary layer regime. From figure 4(a), it is found that the velocity profile of the nanofluid decreases with increase in stretching parameter ($n$) leads to enhance in skin friction coefficient. Figure 4(b) shows that the temperature profile increases with increasing the nonlinear stretching parameter ($n$). Therefore the heat transfer rate reduces with enhancing the value of non-linear stretching parameter ($n$). This signifies that the temperature in the boundary layer region is low for $n=0$ corresponds to the moving sheet with constant ($c$) since the absence of stretching sheet, as compared with the large value of the stretching sheet while opposite trend in the velocity profile as shows in figure 4(a).

\begin{multicols}{2}{\begin{center}
\includegraphics[width=0.93\columnwidth]{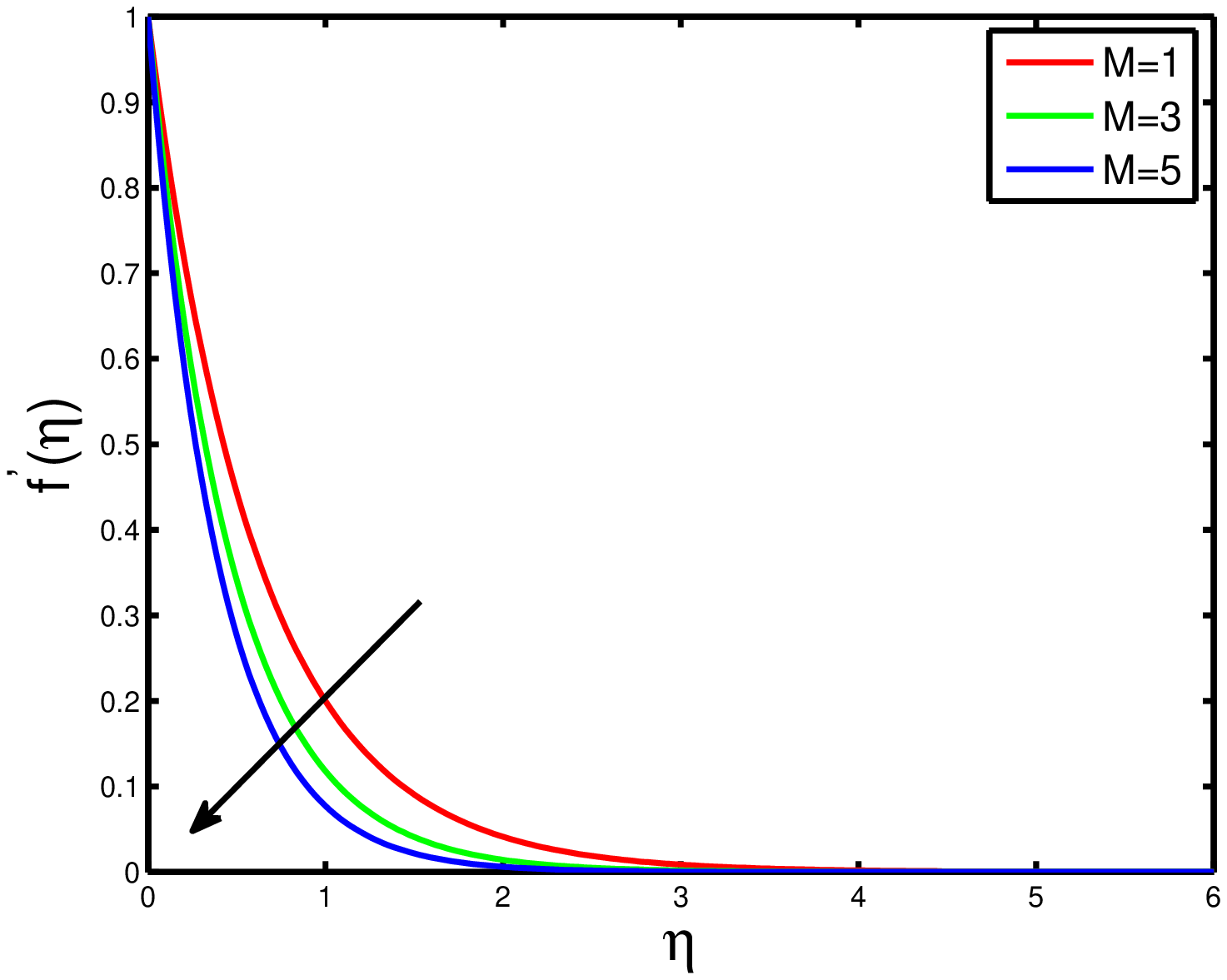}
\end{center}\begin{center}
\textbf{Figure 2(a): Effect on the velocity profile by magnetic parameter $M$.}
\end{center}}
{\begin{center}
\includegraphics[width=0.93\columnwidth]{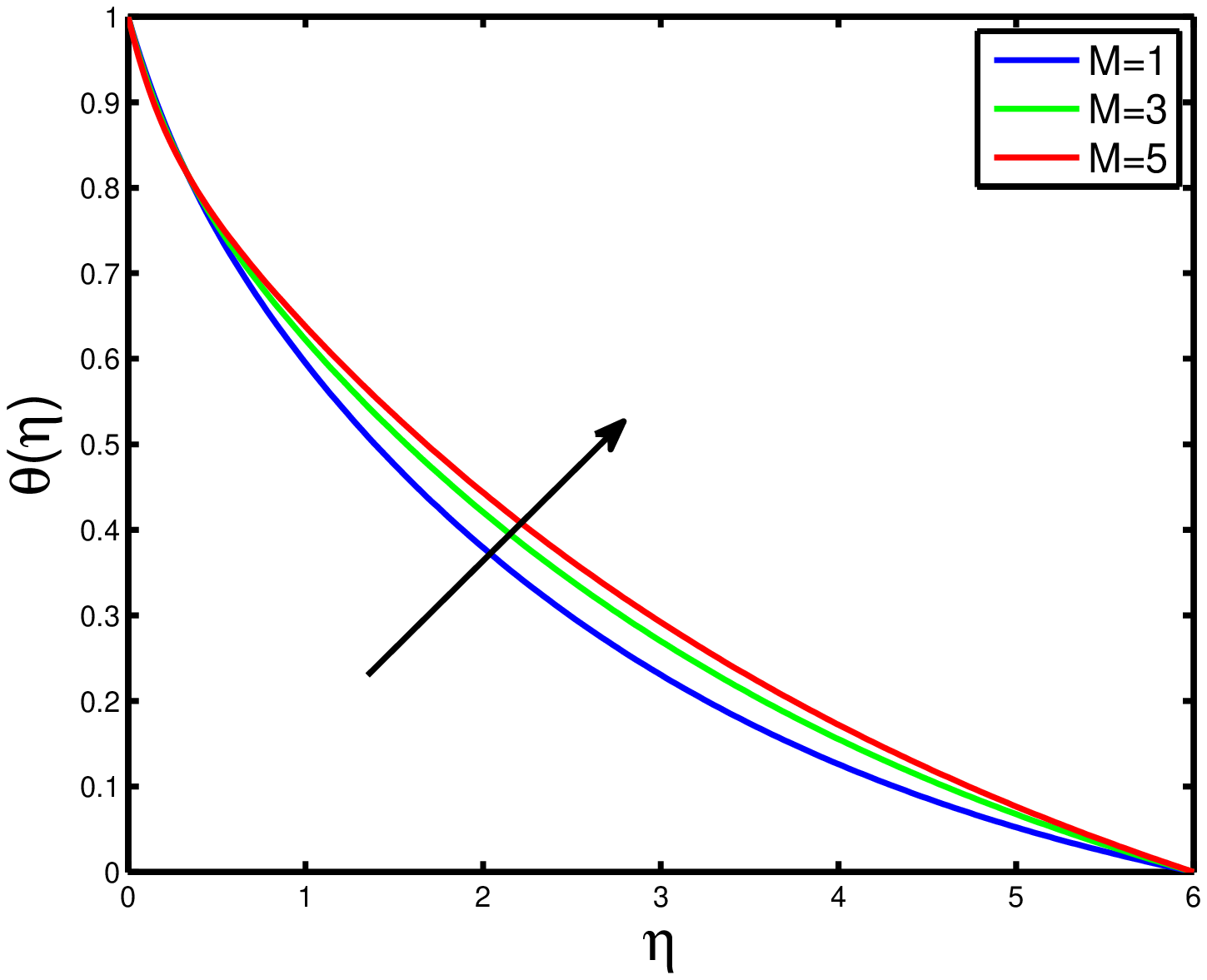}
\end{center}\begin{center}
\textbf{Figure 2(b): Effect on the temperature profile by magnetic parameter $M$.}
\end{center}}
\end{multicols}
\begin{multicols}{2}{\begin{center}
\includegraphics[width=0.93\columnwidth]{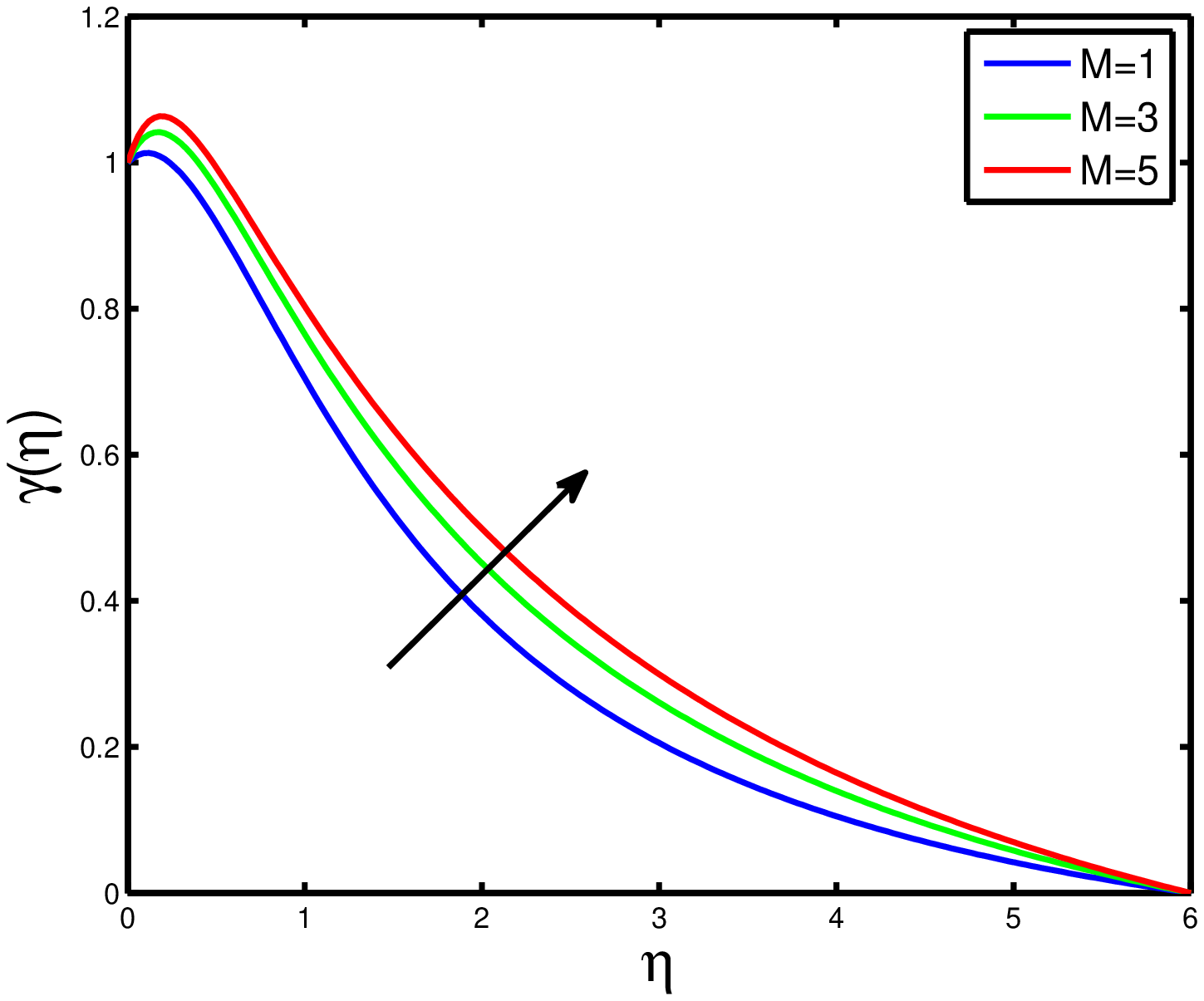}
\end{center}\begin{center}
\textbf{Figure 2(c): Effect on nanoparticle concentration profile by magnetic parameter $M$.}
\end{center}}
{\begin{center}
\includegraphics[width=0.93\columnwidth]{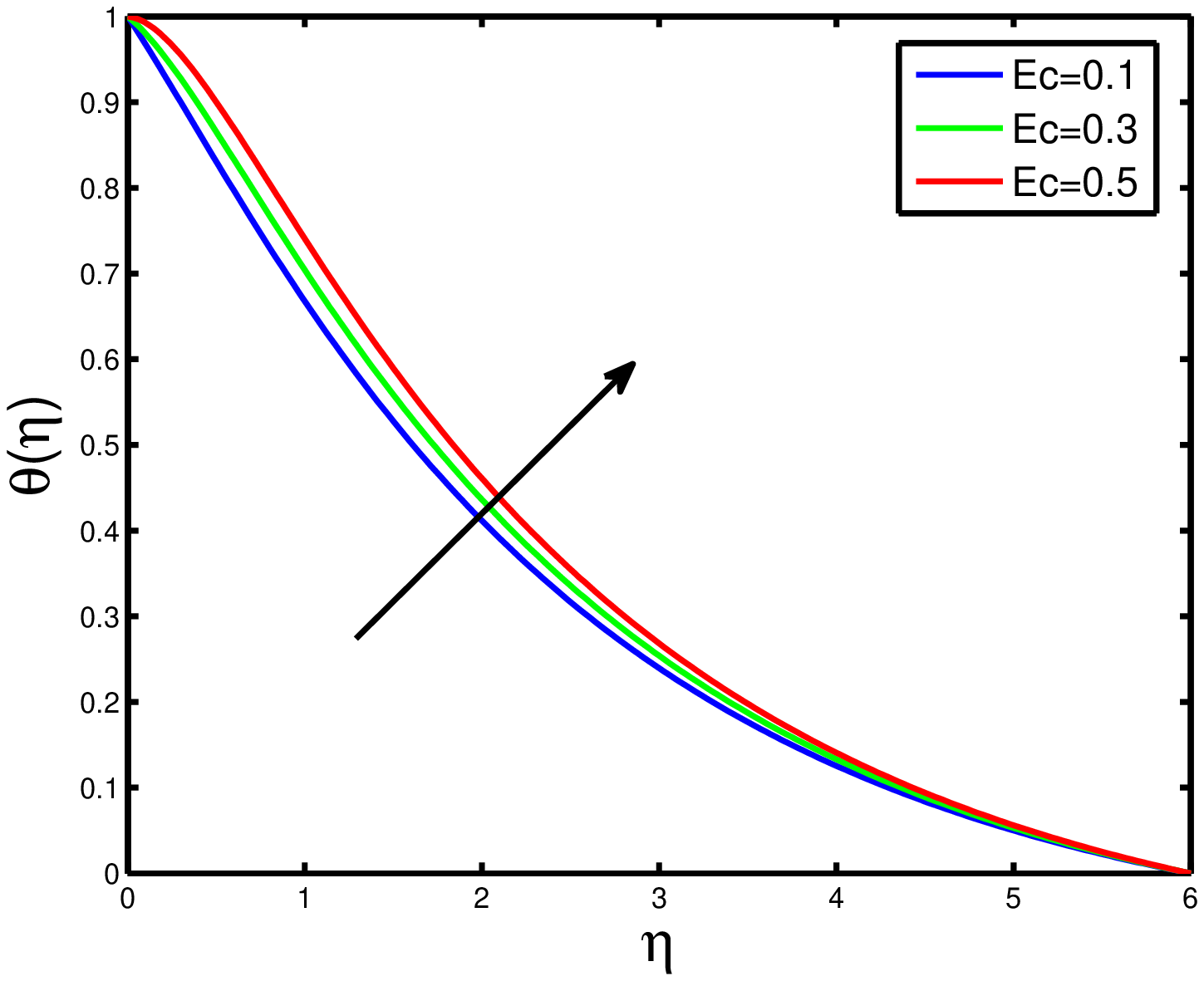}
\end{center}\begin{center}
\textbf{Figure 3: Effect on the temperature profile by Eckert number $Ec$.}
\end{center}}
\end{multicols}
\begin{multicols}{2}{\begin{center}
\includegraphics[width=0.95\columnwidth]{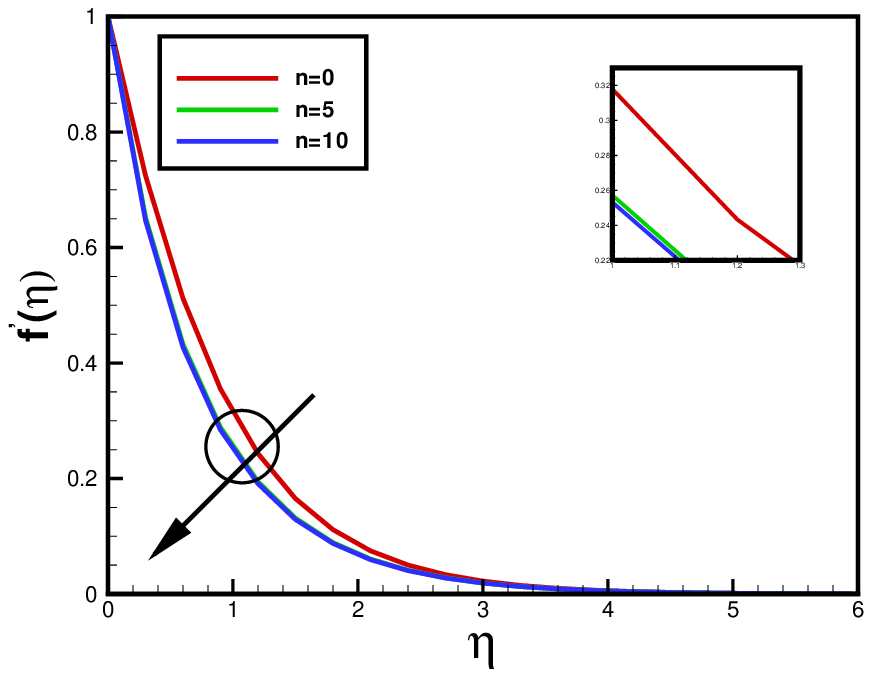}
\end{center}\begin{center}
\textbf{Figure 4(a): Effect on the velocity profile by stretching sheet parameter $n$.}
\end{center}}
{\begin{center}
\includegraphics[width=0.95\columnwidth]{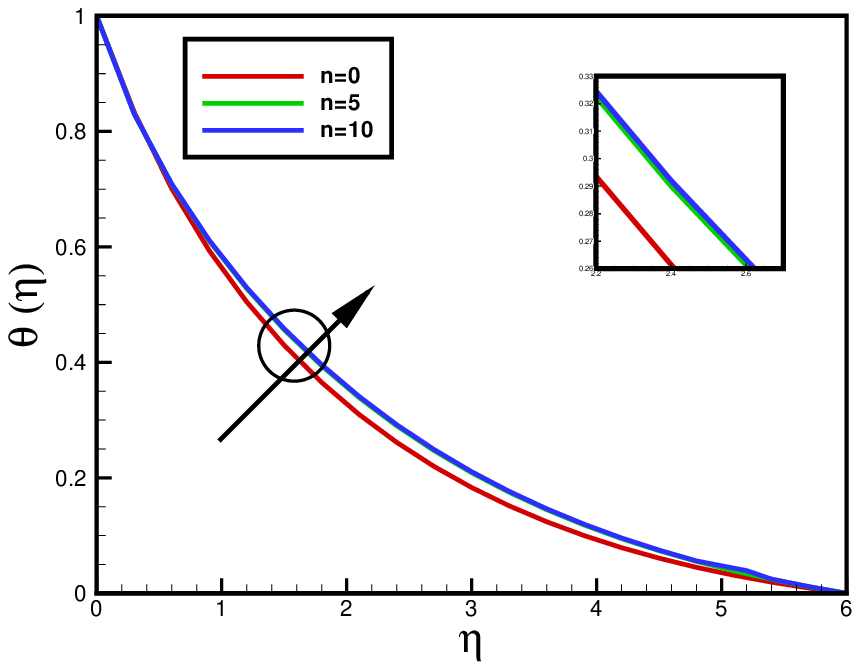}
\end{center}\begin{center}
\textbf{Figure 4(b): Effect on the temperature profile by stretching sheet parameter $n$.}
\end{center}}
\end{multicols}
Figures 5(a) and 5(b) demonstrate the effect of Brownian motion parameter ($Nb$) on the temperature and concentration profiles. As expected that the thermal boundary layer profiles are essentially the same as in the case of regular heat transfer fluids. It is observed that the thermal boundary layer increases as Brownian motion parameter ($Nb$) increases, but the nanoparticle volume fraction decreases with increases in the Brownian motion parameter ($Nb$). On the other hand, the nanoparticle concentration gradient is controlled passively at the sheet by the product of temperature gradient and $\left(\frac{Nt}{Nb}\right)$, when the temperature gradient is fixed, the nanoparticle volume fraction decreases as Brownian parameter ($Nb$) increases because in the fluid motion, the particle deposition away from the fluid regime.\\
Figures 6(a) and 6(b) illustrate the effect of thermophoresis parameter ($Nt$) on the temperature profile and nanoparticle concentration profile. Both temperature and nanoparticle concentration in the boundary layer increases with the increase in the thermophoresis parameter ($Nt$). Consequently, thermophoretic force generated by the temperature gradient which creates a fast flow away from the stretching surface. Accordingly more heated fluid moves away from the stretching sheet, and hence, the temperature in the boundary layer thickness increases with increasing value of $Nt$. Since the thermophoretic force is carried by the fast flow from the stretching sheet which causes an increase in the concentration boundary layer thickness.\\
Figures 7(a), 7(b), 8(a) and 8(b) demonstrate the effects of permeability parameter ($R$) on velocity, temperature, solutal concentration and nanoparticle concentration profiles. It is found that the velocity profile decreases as permeability parameter ($R$) increases, it is shown in figure 7(a). But, increase in the permeability parameter ($R$) leads to increase in all fluid temperature, solutal concentration and nanoparticle concentration. This is due to the fact that the Darcian body force transfer heat from solid surface to the fluid layers.\\
\begin{multicols}{2}{\begin{center}
\includegraphics[width=0.95\columnwidth]{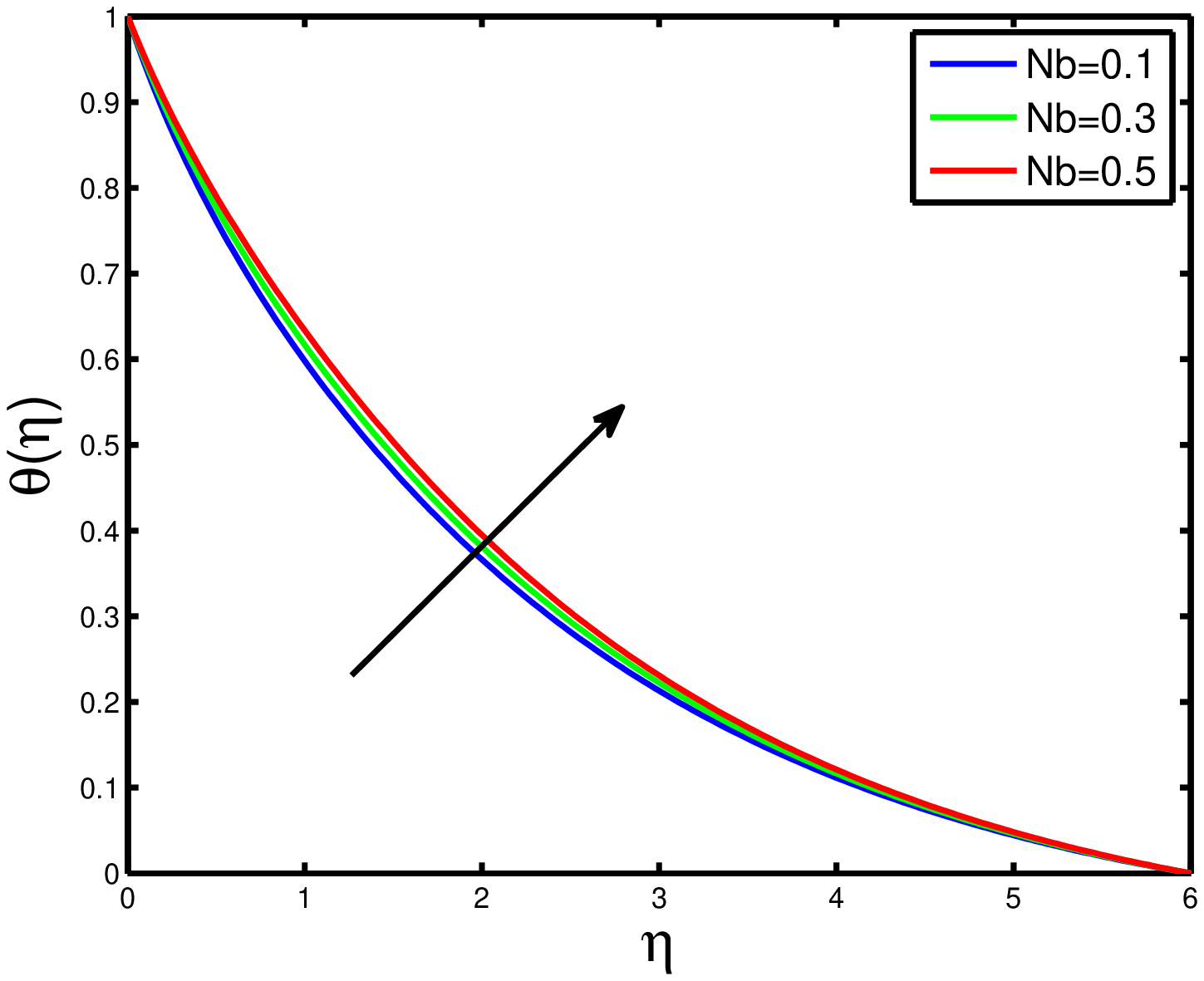}
\end{center}\begin{center}
\textbf{Figure 5(a): Effect on the temperature profile by Brownian motion parameter $Nb$.}
\end{center}}
{\begin{center}
\includegraphics[width=0.95\columnwidth]{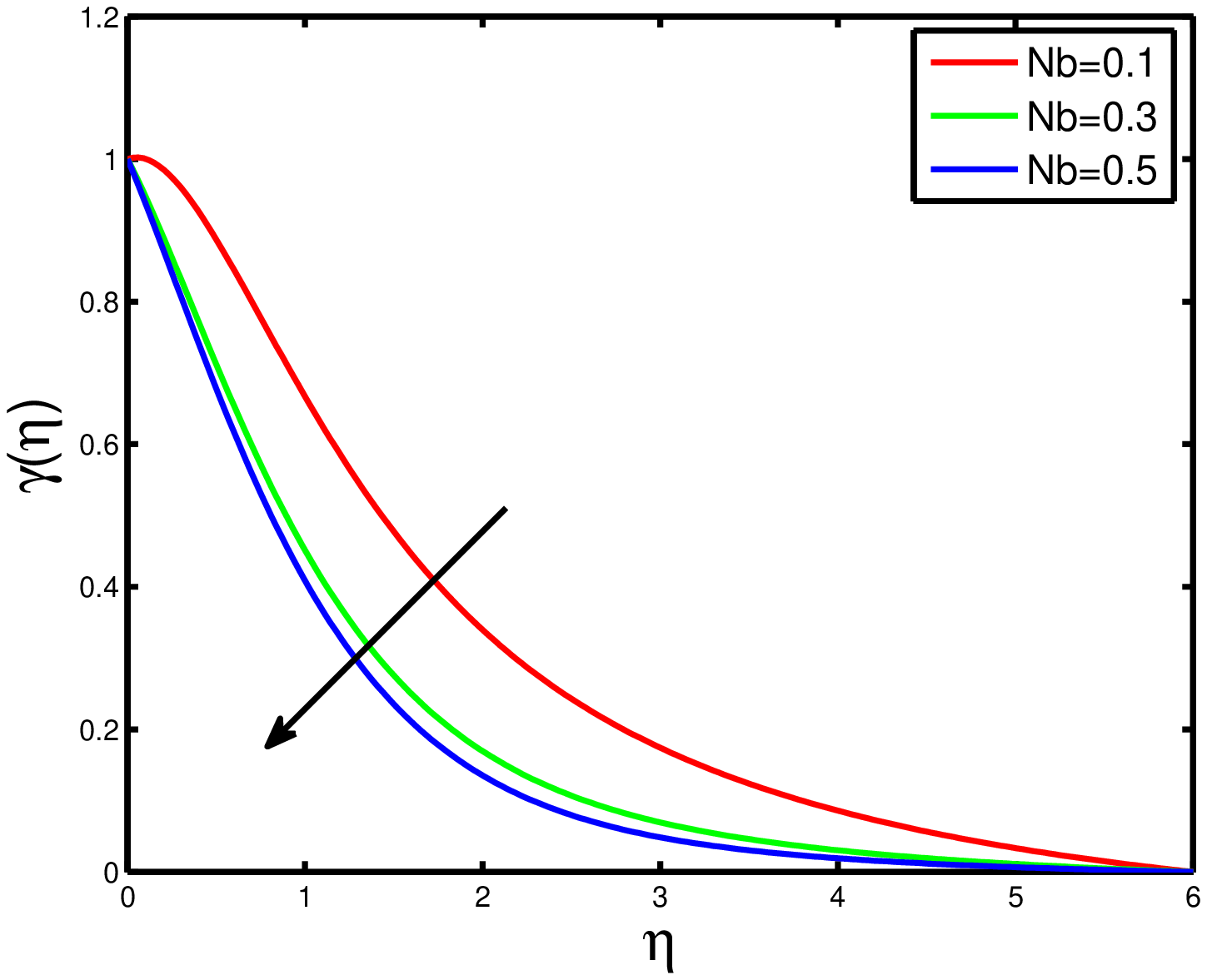}
\end{center}\begin{center}
\textbf{Figure 5(b): Effect on the nanoparticle concentration profile by Brownian motion parameter $Nb$.}
\end{center}}
\end{multicols}

\begin{multicols}{2}{\begin{center}
\includegraphics[width=0.95\columnwidth]{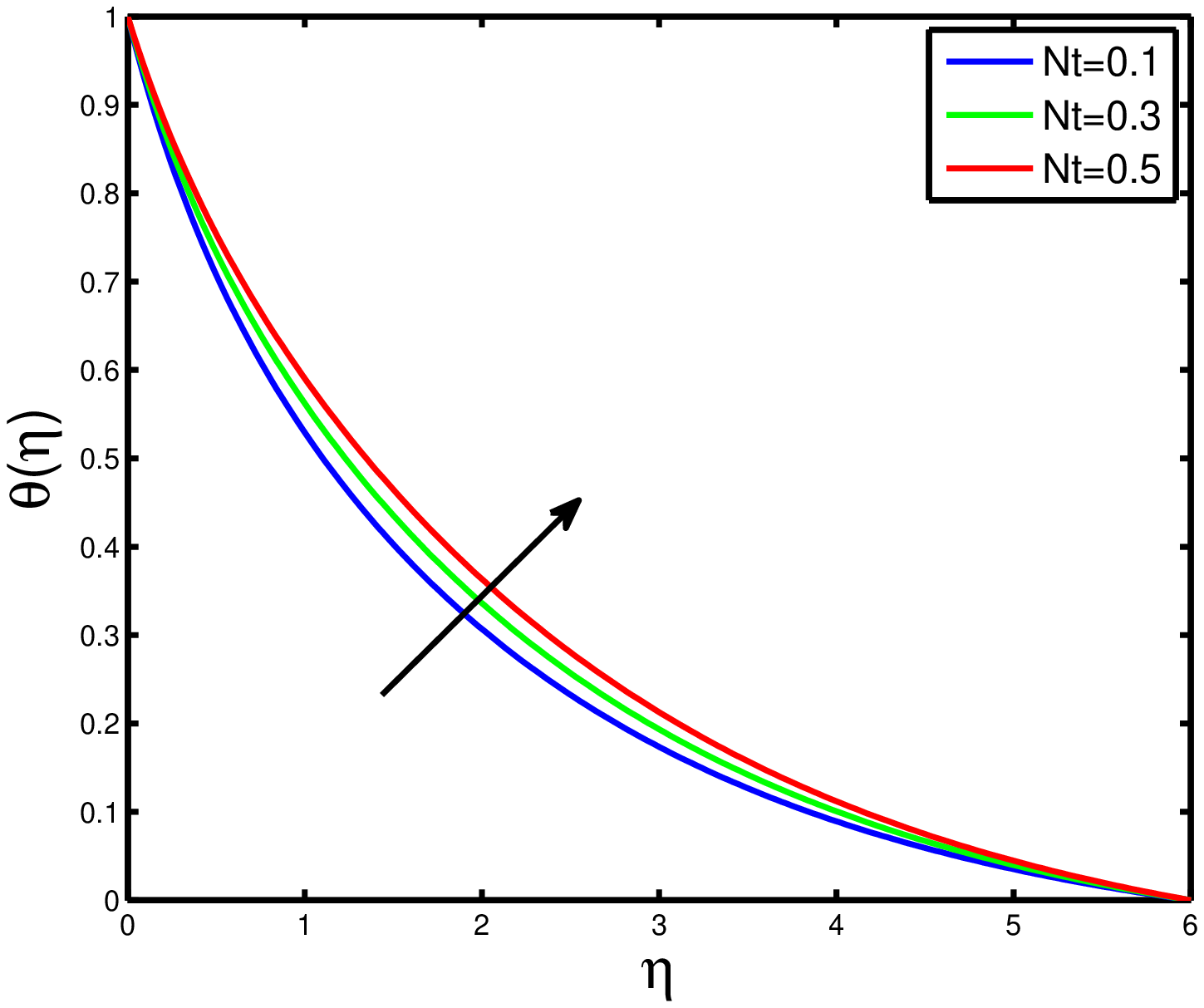}
\end{center}\begin{center}
\textbf{Figure 6(a): Effect on the temperature profile by thermophoresis parameter $Nt$.}
\end{center}}
{\begin{center}
\includegraphics[width=0.95\columnwidth]{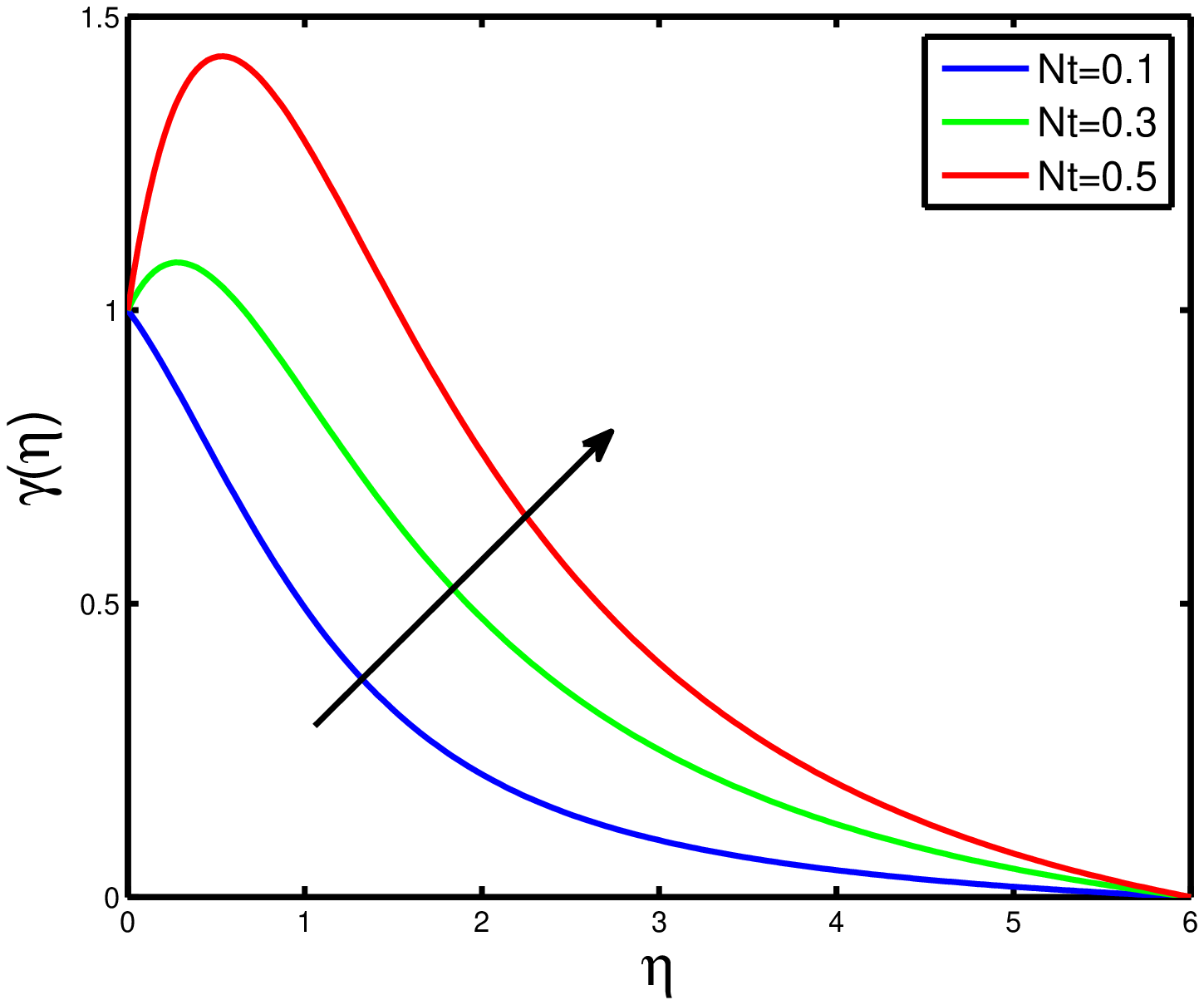}
\end{center}\begin{center}
\textbf{Figure 6(b): Effect on the nanoparticle concentration profile by thermophoresis parameter $Nt$.}
\end{center}}
\end{multicols}

\begin{multicols}{2}{\begin{center}
\includegraphics[width=0.93\columnwidth]{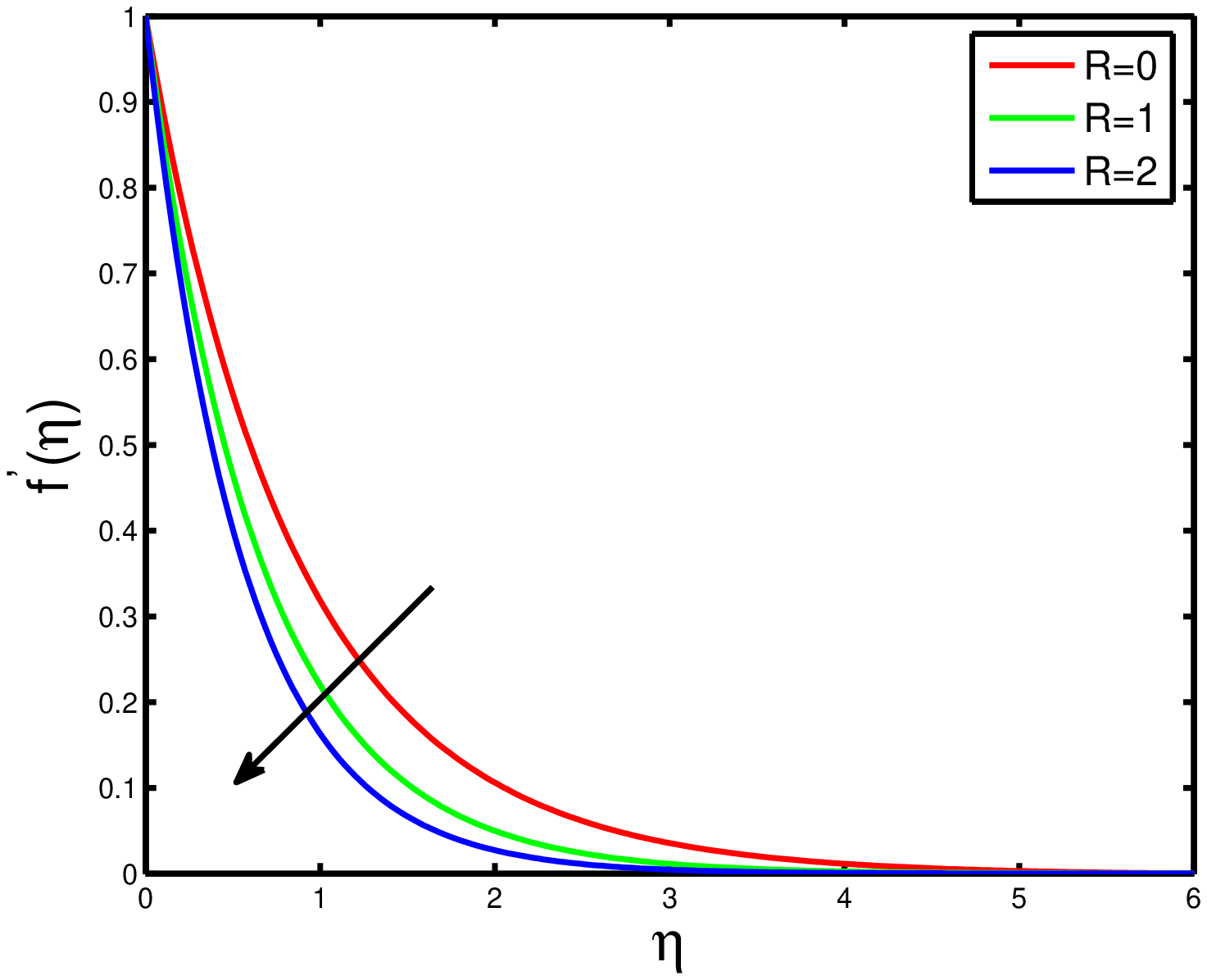}
\end{center}\begin{center}
\textbf{Figure 7(a): Effect on the velocity profile by Permeability parameter $R$.}
\end{center}}
{\begin{center}
\includegraphics[width=0.93\columnwidth]{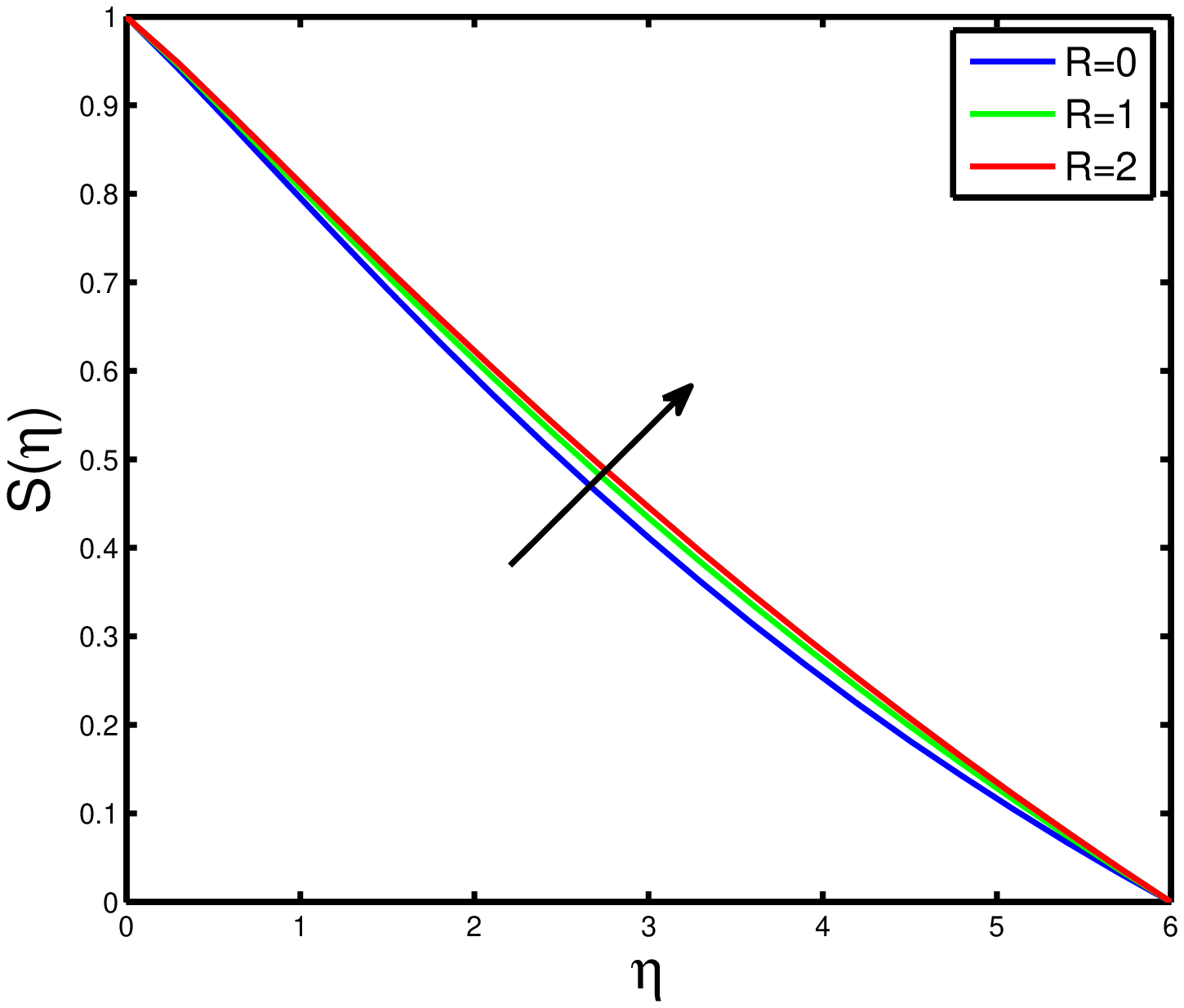}
\end{center}\begin{center}
\textbf{Figure 7(b): Effect on the concentration of salt by Permeability parameter $R$.}
\end{center}}
\end{multicols}
\begin{multicols}{2}{\begin{center}
\includegraphics[width=0.93\columnwidth]{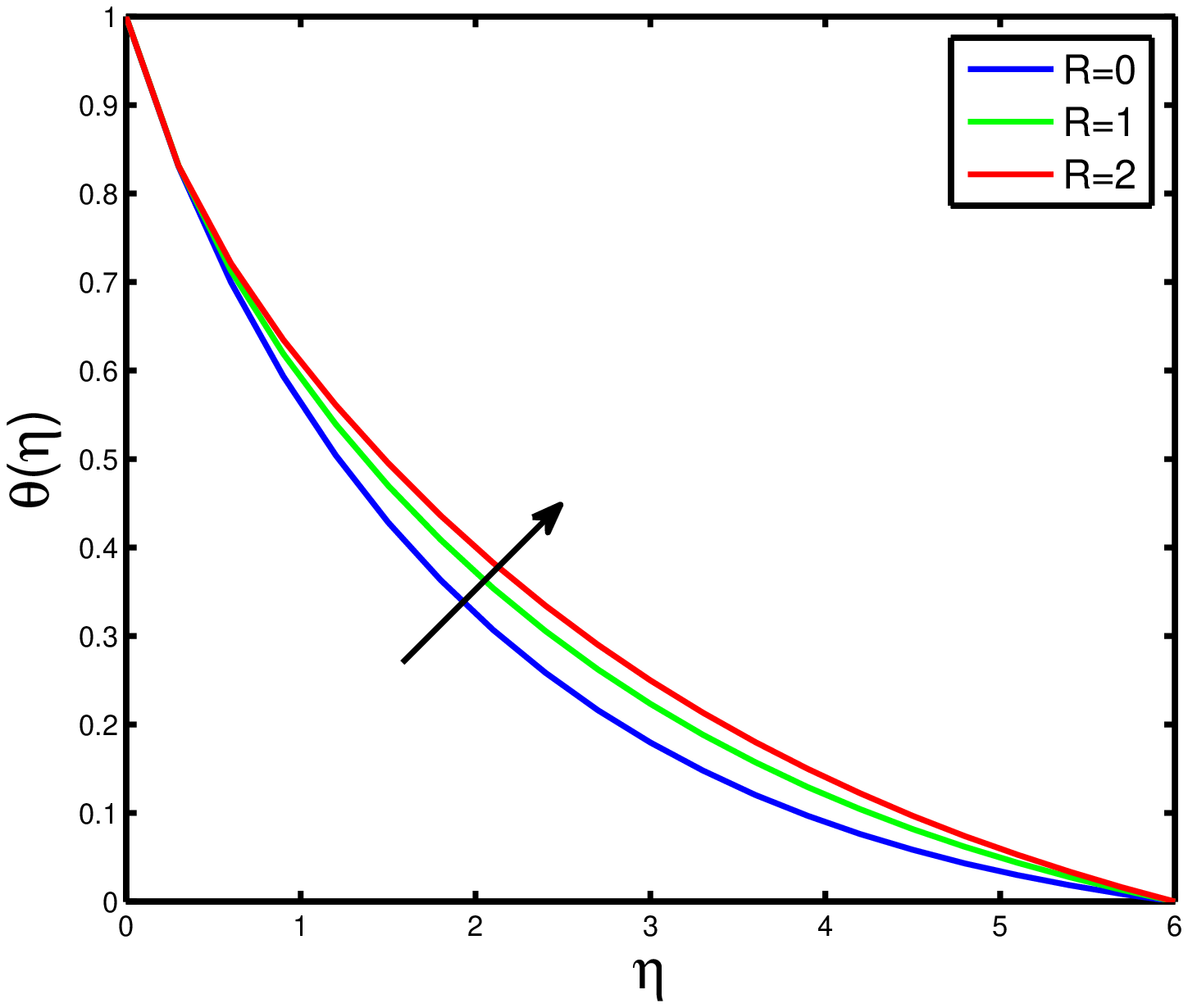}
\end{center}\begin{center}
\textbf{Figure 8(a): Effect on the temperature profile by Permeability parameter $R$.}
\end{center}}
{\begin{center}
\includegraphics[width=0.93\columnwidth]{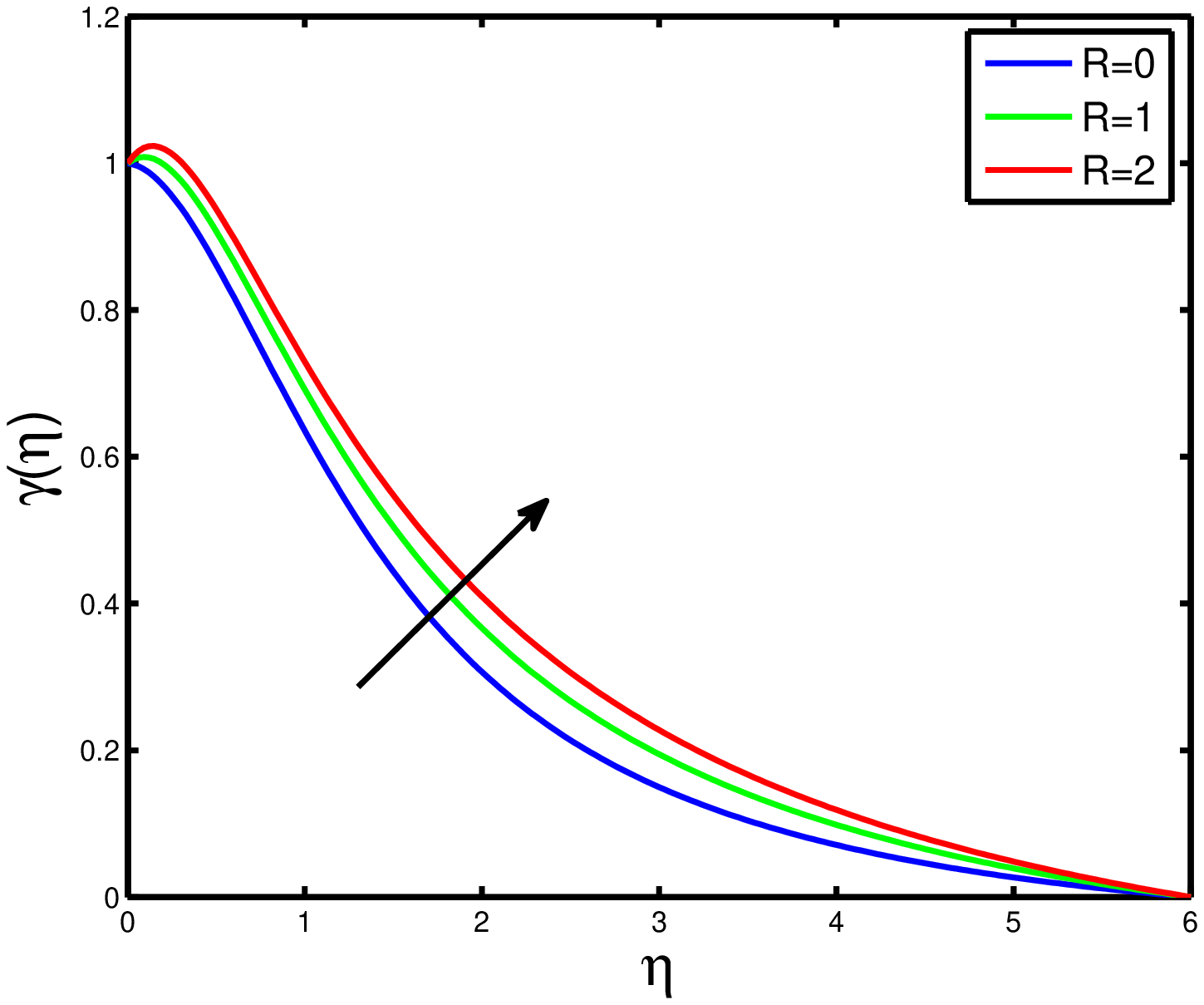}
\end{center}\begin{center}
\textbf{Figure 8(b): Effect on the nanoparticle concentration profile by Permeability parameter $R$.}
\end{center}}
\end{multicols}
Figures 9(a) and 9(b) illustrate the effects of the modified Dufour number ($Nd$) and Dufour-solutal Lewis number ($Ld$) on temperature and nanoparticle concentration profiles respectively. It is observed from figure 9(a) that the temperature profile of the fluid decreases as the value of the modified dufour number increases. It is also noticed that increasing the value of Dufour-solutal Lewis number $Ld$ increases the nanoparticle concentration in the boundary layer regime.\\
Figures 10(a) and 10(b) show the effects of magnetic parameter ($M$), Eckert number ($Ec$) and nanoparticle Lewis number ($Ln$) on the reduced Nusselt number ($Nur$) and nanoparticle Sherwood number ($Shrn$). From both the figures 10(a) and 10(b), It is noticed that increasing values of the magnetic parameter ($M$) decreases the value of both reduced Nusselt number ($Nur$) and nanoparticle Sherwood number ($Shrn$). It is also observed that the reduced Nusselt number decreases with increasing the values of both Eckert number ($Ec$) and nanoparticle Lewis number ($Ln$) because Lewis number is defines the ratio of the thermal diffusivity to the mass diffusivity. Nanoparticle Lewis number is used to characteristic of fluid flows where there is simultaneous heat and mass transfer by convection. On the other hand, Lewis number is also defines the ratio of the Schmidt number to the Prandtl number.  But, the Sherwood number increases as increase the value of $Ec$ and $Ln$. Finally, we observed that the heat transfer rate decreases with all parameters $M$, $Ec$ and $Ln$.
\begin{multicols}{2}{\begin{center}
\includegraphics[width=0.93\columnwidth]{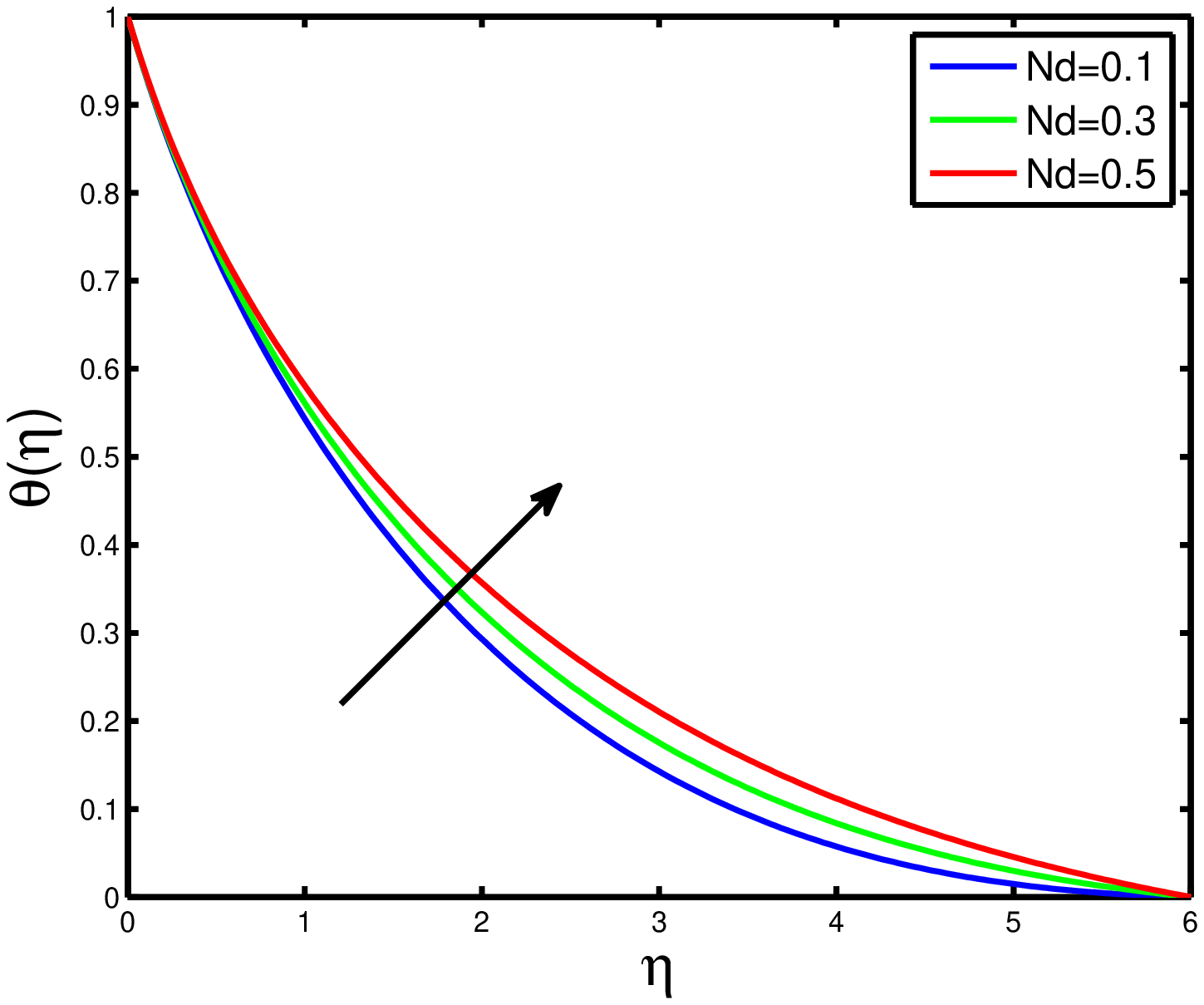}
\end{center}\begin{center}
\textbf{Figure 9(a): Effect on the temperature profile by modified Dufour number of salt $Nd$.}
\end{center}}
{\begin{center}
\includegraphics[width=0.93\columnwidth]{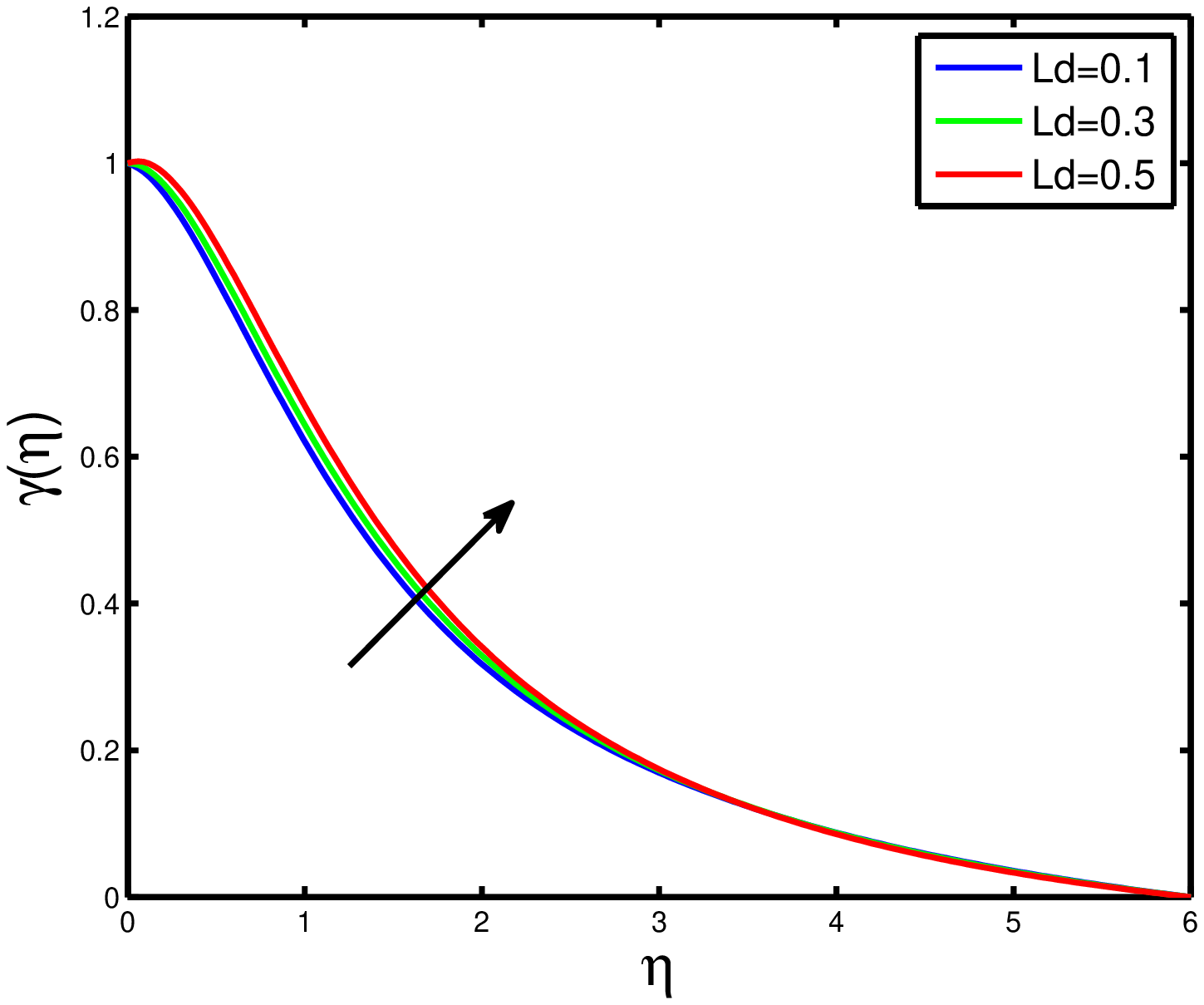}
\end{center}\begin{center}
\textbf{Figure 9(b): Effect on the nanoparticle Sherwood number by Dufour-solutal Lewis number of salt $Ld$.}
\end{center}}
\end{multicols}

\begin{multicols}{2}{\begin{center}
\includegraphics[width=1.0\columnwidth]{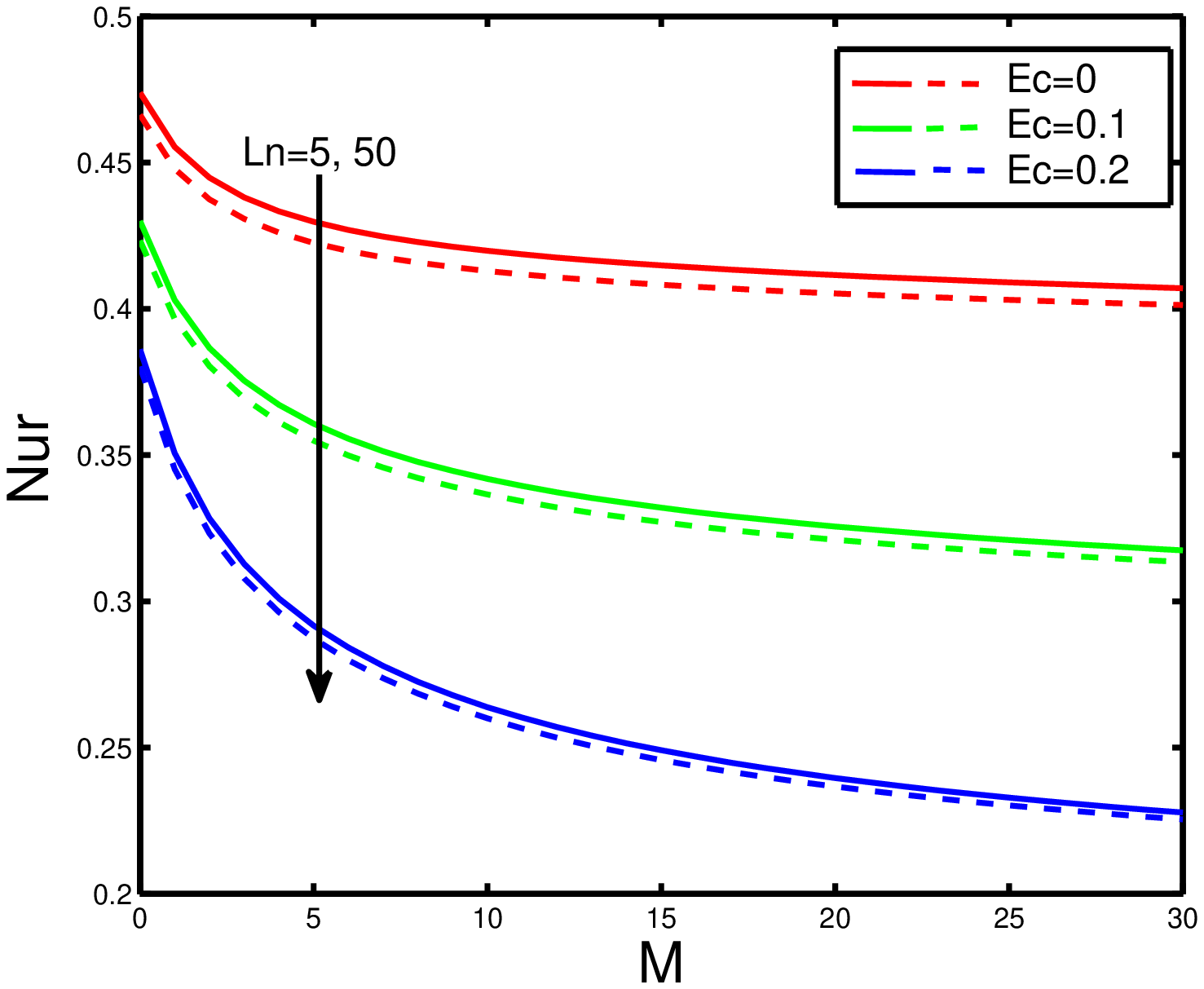}
\end{center}\begin{center}
\textbf{Figure 10(a): Effect on the Reduced Nusselt number ($Nur$) by Magnetic parameter ($M$), Eckert number ($Ec$) and nanoparticle Lewis number parameter ($Ln$).}
\end{center}}
{\begin{center}
\includegraphics[width=1.0\columnwidth]{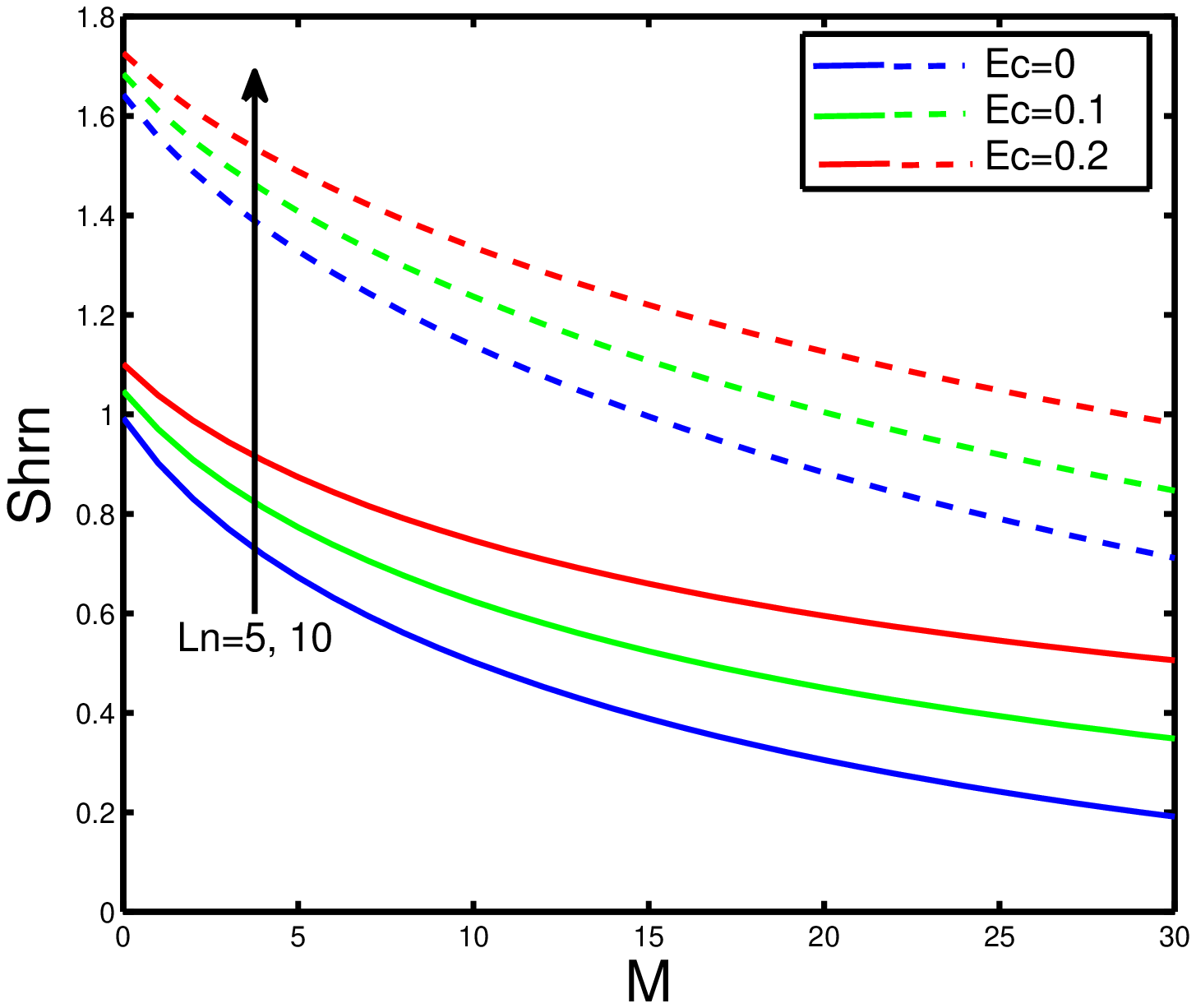}
\end{center}\begin{center}
\textbf{Figure 10(b): Effect on the Sherwood number ($Shrn$) by Magnetic parameter ($M$), Eckert number ($Ec$) and nanoparticle Lewis number parameter ($Ln$).}
\end{center}}
\end{multicols}
\newpage
\begin{multicols}{2}{\begin{center}
\includegraphics[width=1.0\columnwidth]{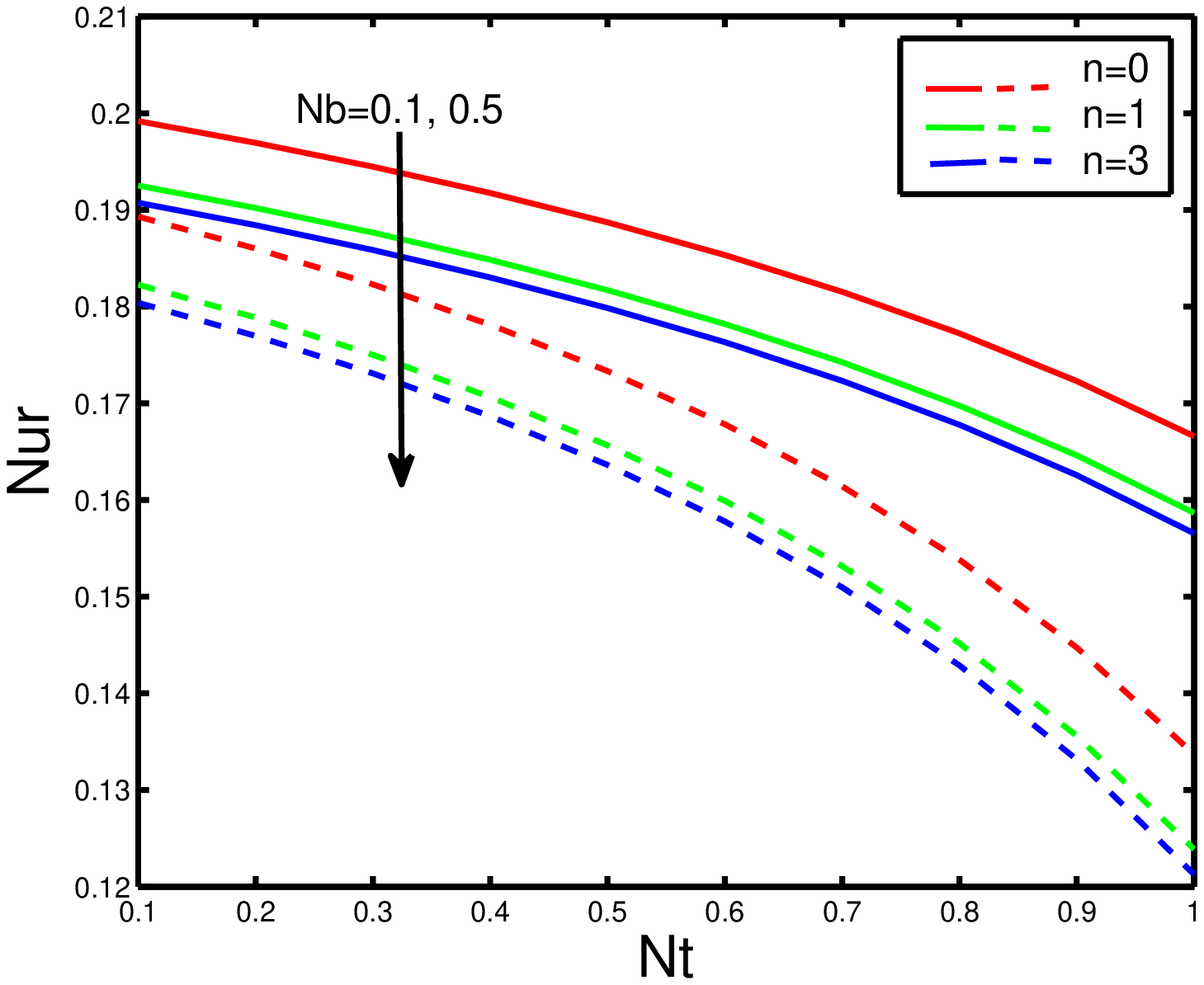}
\end{center}\begin{center}
\textbf{Figure 11(a): Effect on the Reduced Nusselt number ($Nur$) by Brownian motion parameter ($Nb$), thermophoresis parameter ($Nt$) and stretching sheet parameter ($n$).}
\end{center}}
{\begin{center}
\includegraphics[width=1.0\columnwidth]{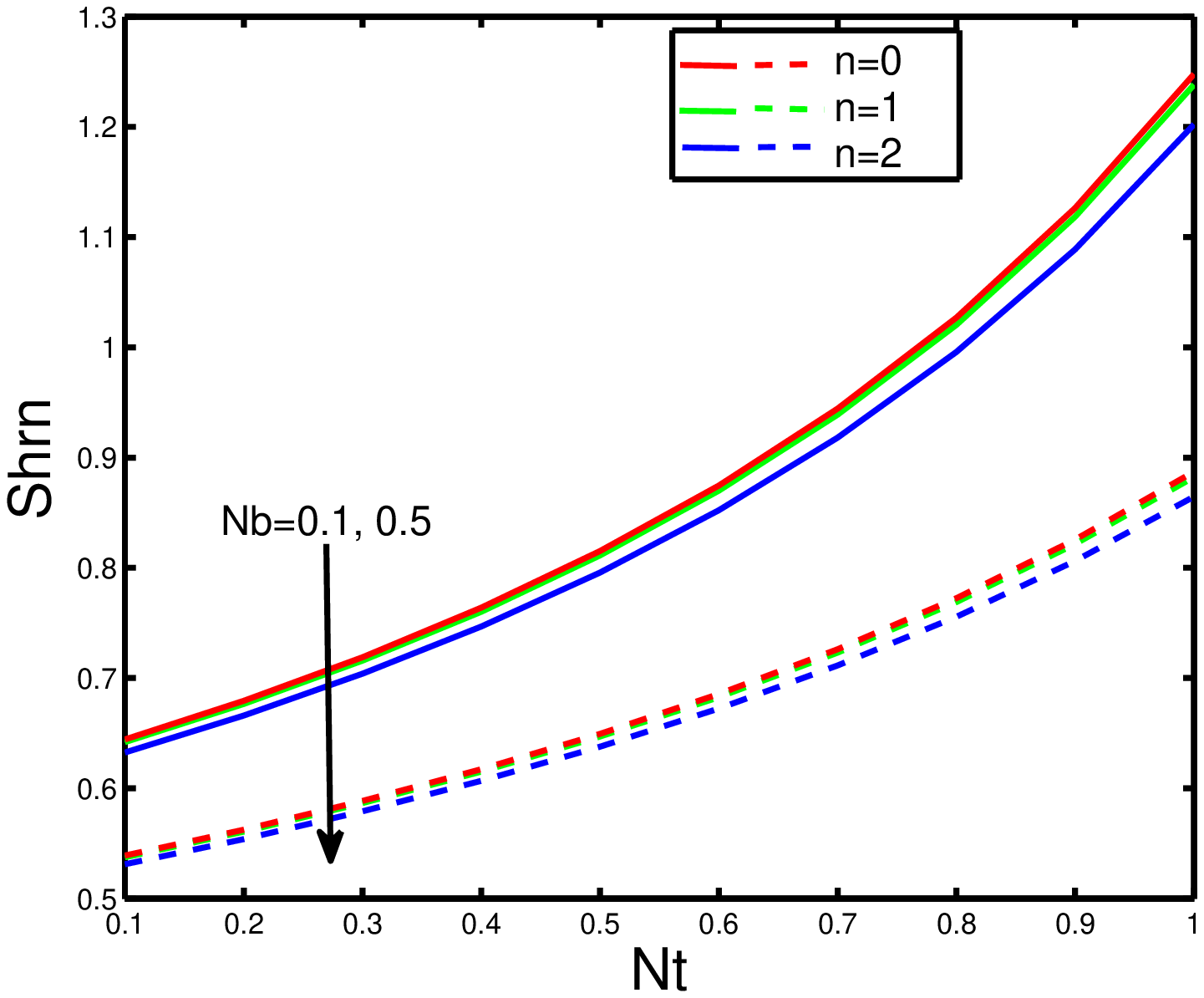}
\end{center}\begin{center}
\textbf{Figure 11(b): Effect on the Sherwood number ($Shrn$) by Brownian motion parameter ($Nb$), thermophoresis parameter ($Nt$) and stretching sheet parameter ($n$).}
\end{center}}
\end{multicols}
Figures 11(a) and 11(b) depict the effect of Brownian motion ($Nb$), thermophoresis ($Nt$) and nonlinear stretching sheet parameter ($n$) on reduced Nusselt number and Sherwood number ($Shrn$). According to figures 11(a) and 11(b), it is evident that both the reduced nusselt number and Sherwood number decreases with increasing the values of Brownian motion parameter ($Nb$) and nonlinear stretching parameter ($n$). It is also found that increasing values of the thermophoresis parameter ($Nt$) decreases the value of the reduced Nusselt number ($Nur$). Where as increasing the value of the thermophoresis parameter ($Nt$) increases the value of the Sherwood number ($Shrn$).

\section{Conclusion}
The triple diffusive MHD boundary layer flow over a nonlinear stretching sheet embedded in a porous medium is solved by Numerically. The conservative form of mass, momentum, energy, solutal and nanoparticle concentration are converted into nonlinear ordinary differential equations by using similarity transformation. A hybrid numerical technique has been used to obtain the solution of these equations. In this study we have show the effect of different parameters on the velocity, temperature, concentrations of both solutal and nanoparticle along with reduced Nusselt number ($Nur$), Sherwood number ($Shr$) and nanoparticle Sherwood number ($Shrn$) and results can be summarized as follows:
\begin{enumerate}
\item Increasing the magnetic parameter ($M$) and nonlinear stretching parameter ($n$) decreases the velocity boundary layer thickness.
\item The temperature profile increases with the increase of the magnetic parameter ($M$) and Eckert number ($Ec$). On the other hand, the effect on the temperature profile is to enhance by magnetic parameter with viscous dissipation.
\item The influence of nonlinear parameter ($n$) is to enhance the heat transfer rate and it has application where fast cooling is required.
\item The heat transfer rate reduces with enhancing magnetic parameter ($M$), thus magnetic field may be used for controlling the cooling.
\item The influences of the Brownian motion parameter ($Nb$) and thermophoresis parameter ($Nt$) is to reduce the heat transfer rate.
\item The permeability parameter ($R$) is reduced to the velocity profile, but increases the temperature, solutal concentration and nanoparticle concentration.
\end{enumerate}
\textbf{Acknowledgements:}
Authors gratefully acknowledge for the financial support from SERB, New Delhi sponsored through a project.

\end{document}